\documentclass[12pt]{iopart}
\usepackage{iopams}
\expandafter\let\csname equation*\endcsname\relax
\expandafter\let\csname endequation*\endcsname\relax
\usepackage{amsmath,graphicx,multirow,xcolor,mathtools}
\usepackage[bookmarks, colorlinks, breaklinks, pdftitle={Impurity scattering in Weyl Semimetals},pdfauthor={Zhoushen Huang}]{hyperref}
\hypersetup{linkcolor=blue,citecolor=blue,filecolor=black,urlcolor=blue}
\usepackage[font=scriptsize,format=hang]{subcaption}
\usepackage{amsfonts}
\begin{document}
\title[Impurity scattering in Weyl Semimetals]{Impurity scattering in
  Weyl semimetals and their stability classification}

\author{Zhoushen Huang$^1$, Daniel P. Arovas$^1$, Alexander V.
  Balatsky$^{2,3,4}$} 

\address{$^1$ Department of Physics, University of California, San
  Diego, La Jolla, CA 92093, USA}

\address{$^2$ Theoretical Division, Los Alamos National Laboratory, Los Alamos,
  NM, 87545, USA}

\address{$^3$ Center for Integrated Nanotechnologies, Los Alamos National Laboratory,
  Los Alamos, NM 87545, USA}

\address{$^4$ Nordic Institute for Theoretical Physics (NORDITA),
  Roslagstullsbacken 23, S-106 91 Stockholm, Sweden}

\ead{zhohuang@physics.ucsd.edu}

\newcommand{\CI}{\emph{\textsf I}}
\newcommand{\CT}{\emph{\textsf T}}
\newcommand{\mib}[1]{{\boldsymbol{#1}}}
\renewcommand{\vec}{\mib}
\newcommand{\Bk}{{\mib k}}
\newcommand{\Bp}{{\mib p}}
\newcommand{\Br}{{\mib r}}
\renewcommand{\Re}{\textsf{Re}}
\renewcommand{\Im}{\textsf{Im}}
\renewcommand{\det}{\textsf{det}}
\renewcommand{\arg}{\textsf{arg}}
\renewcommand{\Tr}{\textsf{Tr}}
\newcommand{\wt}{\widetilde}
\newcommand{\id}{\mathbb{I}}
\newcommand{\gz}{G^{0}}
\newcommand{\hz}{H^{0}}
\newcommand{\grz}{{\cal G}^{0}}
\newcommand{\gr}{{\cal G}}
\newcommand{\ttr}{{\cal T}}
\newcommand{\gc}{\Gamma^{\rm C}}
\newcommand{\ga}{\Gamma^{\rm A}}
\newcommand{\nd}{^{\vphantom{\dagger}}}
\newcommand{\pp}{\text{P}}
\newcommand{\pr}{^{\prime}}
\newcommand{\dg}{^{\dagger}}
\newcommand{\GSB}{\wt\Gamma}
\newcommand{\BHB}{{\scriptscriptstyle{\rm BHB}}}

\begin{abstract}
  Weyl Semimetals (WS) are a new class of Dirac-type materials
  exhibiting a phase with bulk energy nodes and an associated vanishing density
  of states (DOS). We investigate the stability of this nodal DOS
  suppression in the presence of local impurities and consider whether or
  not such a suppression can be lifted by impurity-induced resonances.
  We find that while a scalar (chemical potential type) impurity can
  always induce a resonance at arbitrary energy and hence lift the DOS
  suppression at Dirac/Weyl nodes, for many other impurity types
  (\emph{e.g.} magnetic or orbital-mixing), \emph{resonances are
    forbidden} in a wide range of energy. We investigate a $4$-band
  tight-binding model of WS adapted from a physical heterostructure
  construction due to Burkov, Hook, and Balents \cite{BHB11}, and
  represent a local impurity potential by a strength $g$ as well as a
  matrix structure $\Lambda$. A general framework is developed to
  analyze this resonance dichotomy and make connection with the phase
  shift picture in scattering theory, as well as to determine the
  relation between resonance energy and impurity strength $g$.
  A complete classification of impurities based on $\Lambda$, based
  on their effect on nodal DOS suppression, is tabulated.  We also discuss
  the differences between continuum and lattice approaches.
\end{abstract}
\maketitle

\section{Introduction}
The history of relativistic (Dirac) fermions in solid state band structures
has been known since Wallace \cite{wallace47}, who first considered a single
layer of hexagonal graphite, {\it i.e.\/} graphene.  It was however generally believed
that such structures were intrinsically unstable and impractical to fabricate,
but advances in materials preparation and experimental techniques
have led to a surge of interest \cite{neto09rmp} and a new class of
materials, known as \emph{Dirac materials} \cite{wehling07,HK10,QZ11,WehlingBalatskyUnpub}.
These materials have one or more symmetry-protected Dirac nodes where the
density of state (DOS) vahishes and about which the energy dispersion is linear.
The Dirac nodes are topological in nature, appearing as vortices or monopoles
in the bulk Brillouin zone, and their presence typically requires some fine tuning
such that the appearance of the Dirac structure marks a quantum phase transition
between gapped phases. The properties of the gapped phases on either side of the
transition differ in terms of their boundary behaviors, with the surface or edge
spectra reflecting the topology of the bulk band structure.  Familiar examples of systems
with Dirac nodes include graphene and the class of materials known as topological
insulators \cite{neto09rmp,HK10,QZ11}.

What distinguishes the Weyl semimetals within this framework is that
the nodal structure exists not uniquely at a quantum critical point, but throughout
an entire phase. In $3$D, a Dirac point consists of two Weyl nodes of opposite
chirality overlapping at the same point in $\mib k$-space.  While crystal
symmetry may protect the two Weyl nodes from coupling \cite{young12,wang12},
thereby stablizing the Dirac node, under a general perturbation they will be
coupled and thereby open up a gap. In a system invariant under both time reversal
($\CT$\,) and inversion ($\CI$\,) symmetries, the nodal structure is at least
four-fold degenerate (Kramer's degeneracy), and the aforementioned
separation requires the breaking of either $\CT$ or $\CI$ symmetry, or both.
This can be achieved for example by introducing external field,
magnetic bulk impurities, electron interaction, \emph{etc}.
\cite{BB11,halasz12,go12,sekine12}.  If the perturbation which lifts the
degeneracy is strong enough, a new gapless phase may result, which is the
Weyl semimetal (WS) phase.  The WS has been identified in several recent studies
\cite{Murakami07b,XuPRL11,yang11,wan11,Heikkila12,Krempa12,jiang12},
and is a stable phase because further bulk perturbations can only
shift the Weyl nodes without eliminating them. Recent work has also explored more
exotic band structures involving Weyl fermions, for example their
coexistence with quadratic (massive) fermions
\cite{pickett09,pickett11}, or the symmetry-enforced overlap of Weyl
nodes of the \emph{same} chirality \cite{fang12}. It is known that band
insulators may undergo a topological phase transition as the bulk gap
collapses and re-opens again; in this sense, the WS -- whose bulk gaps
are closed -- is an intermediation of two topologically distinct
gapped phases \cite{Volovik03}, \emph{e.g.} from a trivial insulator
to a topological insulator \cite{Murakami07b}, or a Chern insulator if
$\CT$ is broken \cite{BB11}. 

A minimal model of the WS was constructed by Burkov, Hook, and Balents (BHB)
\cite{BHB11}.  BHB considered a massless $3+1$ dimensional four-component Dirac
fermion model, initially with time-reversal and inversion symmetries.  A $\Bk\cdot\Bp$ expansion
about the $\CT$ and $\CI$\,-symmetric Brillouin zone center yields
the four-component Hamiltonian $\hz(\mib k) = \sum_{a=1}^3
k_a\Gamma^a + m\Gamma^4$, written in terms of Dirac matrices. The sign
of the mass $m$ tells if the bulk insulating state is normal or
topological.  BHB showed that adding a homogeneous $\CT$ and/or $\CI$ breaking term
initially gaps out the Dirac spectrum, but that for sufficiently large symmetry breaking
a WS phase appears, with Weyl nodes occurring at two distinct $\Bk$ points.
(In some cases, the two central bands touch along a circle in $\Bk$-space.)

From the perspective of bulk-boundary correspondence, Weyl nodes may
appear as the ends of the so-called Fermi arc \cite{wan11}, the locus
of gapless surface states interpolating between the projections of
Weyl nodes on the surface Brillouin zone. Such gapless modes
participate in surface transport, with their multiplicity proportional
to the arc length. This gives rise to an anomalous Hall effect (AHE)
of the $\CT$ breaking WS, which recently has been shown to survive
even when the Weyl nodes are subsequently gapped out by node-mixing
scatterings, and is attributed to the persistence of chiral
anomaly \cite{zyuzin12}.

In this paper, we investigate the effects of localized impurities on
the bulk electronic structure of the WS. In particular, we address the
question of whether or not the DOS suppression at Weyl nodes in clean
samples can be lifted via impurity scattering. Local impurities are
modeled as $V = g\Lambda\delta(\mib x)$ where $g \in \mathbb{R}$ is
the coupling strength, $\delta(\mib x)$ restricts the impurity to the
site at $\mib x = 0$, and $\Lambda$ is a matrix structure encoding its
physical type, \emph{e.g.}, $\Lambda = \id$ for scalar (chemical
potential) impurity, and $\Lambda \propto \sigma_z$ for a magnetic
impurity polarized along the $z$ direction. We will speak of the
\emph{stability} of an energy $\omega$ under the scattering of a
$\Lambda$-type impurity in the following sense: if a resonance or
bound state can be induced at $\omega$ by $\Lambda$ with \emph{some}
$g$, then $\omega$ is \emph{unstable} with respect to $\Lambda$.
Otherwise it is stable. Close to the nodal energy, the DOS vanishes as
$\omega^2$. If the nodal energy is unstable, the resulting resonances
will give rise to sharp peaks in the DOS which disrupt the pristine
Dirac spectrum \cite{BSB12}. Bulk transport consequences for scalar
impurities were considered in Refs.
\cite{BHB11,HPV12,halasz12,NHS13,BR13}. Ref. \cite{NHS13} studied the
effect of rare regions in a dirty WS.  Effects of scalar and magnetic
impurities on the surface Dirac nodes of 3D topological insulators
were studied in Ref. \cite{BB10}.

One might be tempted to draw intuition from the more familiar
single-band problems and conclude that an impurity can induce
resonance or bound state at arbitrary energy, given the freedom in choosing
its strength $g$, making all energies unstable.  We find that in the
multiple-band case such as the WS, while this still holds for scalar
impurities, it is not true in general. Instead, stability depends
crucially on the type of impurity, which is mathematically classified
by its commutation relation with the $\Gamma$ matrices in the local
Green's function. For some impurities, resonances and bound states are
forbidden over a wide range of energies. Typically, an impurity is a
foreign atom or local crystalline defect in an otherwise pristine
material. Thus a realistic impurity potential should always involve a
local scalar scattering component.  If this scalar effect dominates the
impurity, intragap resonances can be induced which will destabilize the Weyl node
at a single particle level. If, on the other hand, the scattering is
dominated by the resonance-forbidding components, then the Weyl node will remain.

This paper is organized as follows: In Sec.~\ref{sec:method}, we
present a general framework to address the existence of impurity
resonances, and the dependence of their energies on the impurity
strength. In Sec.~\ref{sec:model}, we introduce a four-band tight
binding lattice model of the WS in terms of the $\Gamma$ matrices,
adapted from the continuum BHB model \cite{BHB11}. Before analyzing
the impurity effect in this lattice model, we first discuss in
Sec.~\ref{sec:cont} the situation in the low energy theory, namely the
original continuum BHB model, and show that a natural momentum cutoff
around the Weyl nodes in such theories dismisses the important physics
of a stabilized Weyl node. We then turn to a full lattice treatment:
In Sec.~\ref{sec:imp}, we apply the method developed in
Sec.~\ref{sec:method} to the WS model and classify impurities
$\Lambda$ according to their effect on the electronic structure. We
shall throughout the paper restrict $\Lambda$ to being one of the
sixteen $\Gamma$ matrices (including $\id$); linear combinations
thereof can be analyzed in the same fashion but with more tedious
algebra. We will first illustrate the method in Sec.~\ref{sec:ti-sym}
with the simpler case of a Dirac semimetal which is invariant under
both $\CT$ and $\CI$. The (fine-tuned) Dirac node is shown to be
unstable with $\CI$ even impurities, but stable with $\CI$ odd ones.
We apply the same approach to WS with a symmetry breaking term $\eta
\GSB$ where $\GSB$ is a matrix and $\eta$ is its strength. For a
generic energy $\omega$, we find that its stability depends on both
the impurity type $\Lambda$ and the symmetry breaking term $\eta\GSB$.
Sec.~\ref{sec:ibreak} discusses $\CI$ breaking WS (which may or may
not break $\CT$) where $\GSB$ anticommutes with $\CI$. In this case,
the Weyl node energy is found to be stable for any $\Lambda$ that does
not commute with the local Green's function, but unstable if it
commutes. Sec.~\ref{sec:tbreak} discusses the $\CI$\,-symmetric WS, in
which $\GSB$ necessarily breaks $\CT$. Again, impurities commuting
with the local Green's function will disrupt the Weyl node stability.
Those that do not fully commute yield either a nodal energy stable
over the full range of $\eta$, or a critical symmetry breaking
amplitude $\eta_c$ -- which is fully determined by parameters of the
clean system -- that splits the $\eta$ axis into two phases where the
nodal energy is stable in one phase and unstable in the other. The
critical amplitude $\eta_c$ is found to be related to a type of band
inversion and indicates a phase transition in the gap of resonant
impurity band structure, reminiscent of band inversions in
topological/Chern insulators that are responsible for topological
phase transitions. We conclude in Sec.~\ref{sec:summary}.

\section{Resonance criteria in a generic multi-band system}
\label{sec:method}

\begin{figure}
  \centering
  \begin{subfigure}[t]{0.4\textwidth}
    \centering
    \includegraphics[width=\textwidth]{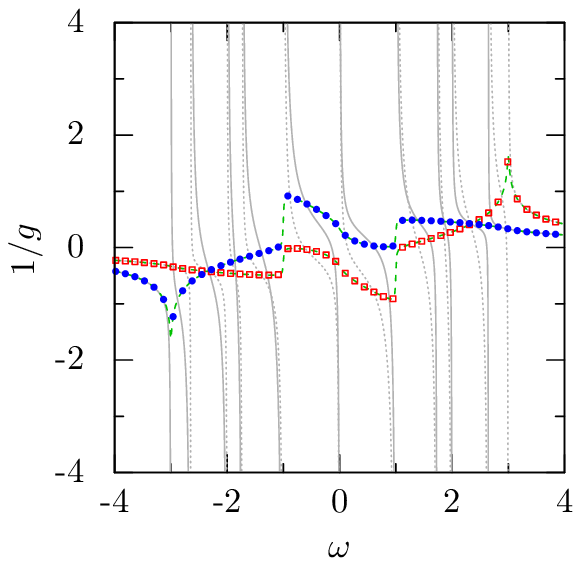}
    \caption{Spectral evolution ($6\times 6$ unit cells)}
    \label{gra-spec-6x6}
  \end{subfigure}
  \begin{subfigure}[t]{0.4\textwidth}
    \centering
    \includegraphics[width=\textwidth]{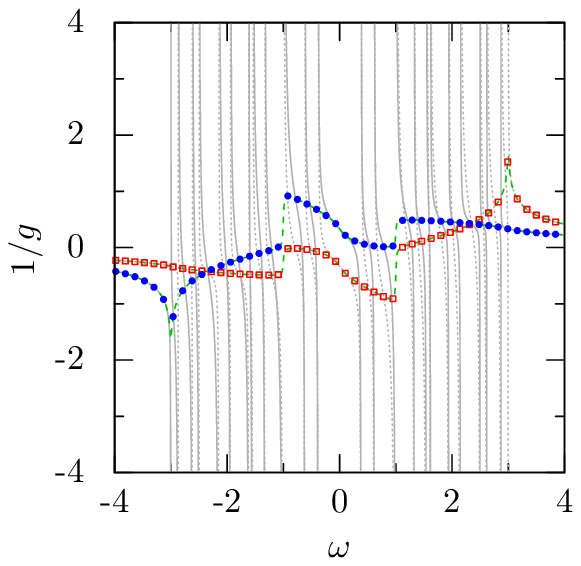}
    \caption{Spectral evolution ($10 \times 10$ unit cells)}
    \label{gra-spec-10x10}
  \end{subfigure}

  \begin{subfigure}[t]{0.4\textwidth}
    \centering
    \includegraphics[width=\textwidth]{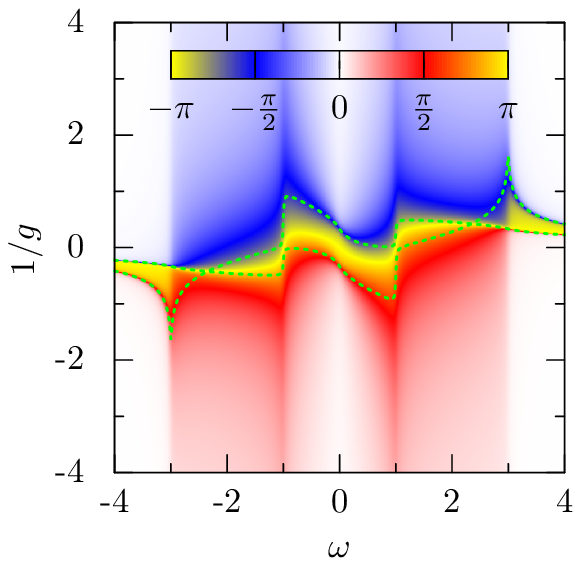}
    \caption{Phase shift}
    \label{gra-phase}
  \end{subfigure}
  \begin{subfigure}[t]{0.4\textwidth}
    \includegraphics[width=\textwidth]{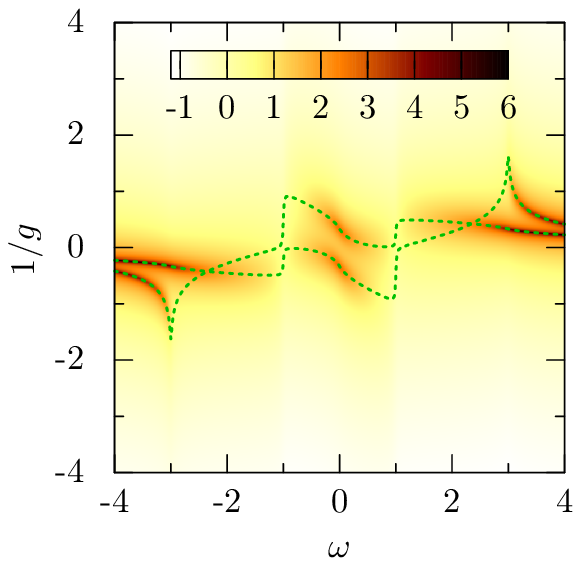}
    \caption{$\log||T(\omega + i0^{+})||$}
    \label{gra-tnorm}
  \end{subfigure}
  \caption[Graphene]{
    (Color online) Various ways to identify bound states and resonances 
    for a scalar impurity $V = g\id$ in graphene.  Clean
    graphene is modeled as $\hz(\mib k) =\left(
      \begin{smallmatrix}
        0 & \gamma_{\mib k} \\ \gamma_{\mib k}^{*} & 0
      \end{smallmatrix}\right)$
    where $\gamma_{\mib k} = 1 + e^{-ik_1} + e^{-ik_2}$ and $k_i =
    \mib k \cdot \mib a_i$, with $\mib a_i$ ($i=1,2$) being the two
    primitive direct lattice vectors. Panels (a) and (b) show the
    spectral evolution of $H = \hz + g\id$ (solid and dotted gray
    curves) with impurity strength $g$, for different lattice sizes.
    These are solved as zeros of the eigenvalues of the $2\times 2$
    matrix $T^{-1}(\omega,g) = g^{-1}\id - \gz_{00}(\omega)$, with the
    eigenvalue index encoded by different line color/types. Migration
    of such levels between different clean states ($g^{-1} = \pm
    \infty$ limit) constitutes a resonance (inside bulk bands) or a
    bound state (outside bulk bands), and can be extracted as zeros of
    $\ttr^{-1}(\omega, g)$, the Hermitian part of $T^{-1}$, shown as
    dashed green curves, with solid circular blue points overlaying on
    the branch corresponding to the solid spectral flow and red empty
    square on the branch corresponding to the dotted spectral flow
    (see text). The discontinuity in the green curves at $\omega = \pm
    1$ is concomitant with the Van Hove singularity in the DOS (not
    plotted) (c) shows the phase shift $\arg\,\det\, T(\omega +
    i0^{+}, g)$ in the thermodynamic limit, where $\pm \frac{\pi}{2}$
    (heaviest red/blue) could be interpreted as resonance or bound
    state, see text. (d) plots the norm of retarded $T$ matrix, which
    can be used to distinguish between resonance and anti-resonance
    that is hard to tell from (c), the former corresponding to the
    dark feature and the latter suppressed in such a plot. The dashed
    green curves in (c) and (d) are the same as those in (a) and (b).}
  \label{graphene}
\end{figure}

The effect of localized impurities can be studied in general using the
standard $T$-matrix formalism \cite{Economou06,balatsky06}. We briefly
summarize the procedure below and establish notation. Given a
Hamiltonian $H = \hz + V$, its Green's function is
\begin{equation}
G(z) \equiv (z -H)^{-1} = \gz + \gz \,T\, \gz\ ,
\end{equation}
where $z\in \mathbb{C}$ is the complex
frequency, $\gz(z) \equiv (z - \hz)^{-1}$ is the Green's function of
$\hz$, and $T = V(\id - \gz V)^{-1}$ is the $T$-matrix. Assume the
impurity potential $V$ is localized at a spatial point $\mib r = 0$:
$V_{\mib r \mib r\pr} = g \Lambda\, \delta_{\mib r,0}\,\delta_{\mib r\pr,0}$,
where $g$ is the potential strength and $\Lambda$ is a dimensionless
matrix whose rank is equal to the number of bands. The Green's
function connecting $\mib r$ and $\mib r\pr$ is $G_{\mib r \mib r\pr}
= \gz_{\mib r \mib r\pr} + \gz_{\mib r 0} \, T\nd_{00} \, \gz_{0 \mib r\pr}$,
and $T_{00}(z) = [g^{-1} \Lambda^{-1} - \gz_{00}(z)]^{-1}$ is the only
nontrivial block of the $T$-matrix. For translationally invariant
systems, the local Green's function is $\gz_{\mib r \mib r} =
\gz_{00}(z) = \frac{1}{N} \sum_{\mib k} (z - \hz_{\mib k})^{-1}$ where
$\hz_{\mib k}$ is the Fourier transform of $\hz$ and $N$ is the number
of $\mib k$ points, {\it i.e.\/} the number of unit cells of the crystal.
The corresponding local density of states (LDOS)
at site $\mib r$ with energy $\omega$ is $\rho_{\mib r}(\omega) = -\Im
\Tr\, G_{\mib r \mib r}(\omega + i0^{+})$.

Bound states and resonances are consequences of the energy spectrum
reconstruction induced by impurities. \emph{Before} taking the
thermodynamic limit, eigenvalues of $H$ are poles of $T(\omega)$ on
the real $\omega$ axis, {\it i.e.\/} the zeros of $T^{-1}(\omega) =
g^{-1}\Lambda^{-1} - \gz_{00}(\omega)$. Upon tuning of the
impurity strength $g^{-1}$, each pole will trace out a curve in the
$(\omega, g^{-1})$ plane. Note that $g^{-1} = \pm \infty$ are
identified as $g = 0$, and $g = \pm \infty$ as $g^{-1} = 0$. Thus one
may consider an adiabatic cycle in which $g^{-1} : -\infty \rightarrow 0
\rightarrow + \infty$. After a full cycle, the poles must collectively
recover their initial positions, which are the set of clean states at
$g^{-1} = \pm \infty$. Individual poles may either stick close to one
clean state, or migrate between different ones. An example is shown in
Figs.~\ref{gra-spec-6x6} and \ref{gra-spec-10x10} for the spectral
evolution of a graphene sheet of $6\times 6$ and $10\times 10$ unit
cells, respectively, in the presence of a scalar impurity.

A point $(\omega, g^{-1})$ on the spectral evolution curves
outside the bands of the clean system represents a bound state
of energy $\omega$ induced by an impurity of strength $g$. For those
inside the clean bands, one has to distinguish between poles very
close to -- hence mere perturbations of -- a clean state, and those in
the middle of a migration. The latter, like bound states, are
manifestations of the impurity effect. They differ in that an increase
in system size $N$ has little influence on the bound states, but will
split spectral lines inside the clean bands to accommodate newly
created clean states; see Fig.~\ref{gra-spec-6x6} and
\ref{gra-spec-10x10} for this lattice size effect. What remains
unchanged when increasing $N$ is the trend of rapid pole migration.

An effective way to extract the locus of such rapid migrations,
which we identify as resonances, is to find the zeros of
$\ttr^{-1}(\omega, g^{-1}) \equiv g^{-1}\Lambda^{-1} -
\grz_{00}(\omega)$, where $\grz_{00}(\omega) \equiv
\frac{1}{2}(\gz_{00}(\omega + i\epsilon) + [\gz_{00}(\omega +
i\epsilon)^{\dagger}])$ is the Hermitian part of the retarded local
Green's function, where the imaginary part $\epsilon$ is taken to be
greater than the energy spacing between consecutive bulk levels. In
so doing, divergences in $T$ originating from poles of
$\gz_{00}(\omega)$ itself are eliminated from the zeros of
$\ttr^{-1}(\omega)$, leaving only those caused by the aforementioned spectral
migrations. The zeros of $\ttr^{-1}(\omega)$ are shown for the case of graphene
as green curves in Figs.~\ref{gra-spec-6x6} and \ref{gra-spec-10x10}.
In these figures, they sit close to the inflection points of the
spectral curves, where one might say the pole migration is most rapid.

In the thermodynamic limit $N \rightarrow \infty$, the Hermitian matrix
$\ttr^{-1}$ remains well defined. Since $\epsilon \rightarrow 0^{+}$,
$\ttr^{-1}$ and $T^{-1}$ are identical outside bulk bands, the
zeros of $\ttr^{-1}$ can be used to identify both bound states and
resonances. Callaway showed \cite{callaway67} that the phase shift at
$\omega$, \emph{viz.} $\delta(\omega) \equiv \arg\, \det\, T(\omega +
i0^{+})$, equals $\pi \times \Delta N(\omega)$ where $\Delta
N(\omega)$ is the difference in the \emph{total} number of states of
$H$ and of $\hz$ below $\omega$. One definition of resonance in this
context is for the phase shift to be $\pm \pi/2$, \emph{viz.}
$\Re\,\det\, T = 0$, the reason being that the number of extra states
is a half-odd-integer, which represents the ``center'' of the
process in which one extra state is gained or lost. The former is
called a resonance and the latter an anti-resonance. Over the
full range of $\omega$ they must balance each other if introduction of
an impurity does not change the total number of states, a form of the
Friedel sum rule. This is related to our resonance criteria in that
the zero of $\ttr^{-1}$ is the center of a spectral migration -- a
spectral line migrating past $\omega$, by definition, contributes
$|\Delta N(\omega)| = 1$ to the total number of states below $\omega$.
The difference between our criteria and the phase shift picture is
that there are in general $N_B$ branches of spectral evolutions at any
$\omega$ where $N_B$ is the number of bands, {\it e.g.\/} $N_B = 2$ in
the graphene example shown in Fig.~\ref{graphene}. Our criteria is
essentially to track the increment/decrement contributed by any single
branch, whereas $\Re\,\det\,T = 0$ takes into account all branches. In
any case, the difference is consequential only in identifying the
location of the anti-resonances. For comparison, we plot the phase shift
(color map) together with the zeros of $\ttr^{-1}$ (dotted green line)
for a graphene impurity in Fig.~\ref{gra-phase}.

The distinction between resonance and anti-resonance is not immediately apparent
from the phase shift plot. It relies on the
sign of $s = \partial \delta/\partial \omega$: $s > 0$ is a resonance
and $s < 0$ an anti-resonance. Furthermore, in cases where two bound
states/resonances are close together -- or even degenerate as can
happen in Dirac semimetals to be discussed in Sec.~\ref{sec:ti-sym}
-- the phase shift will experience a $2\pi$ change over a small
$\omega$ window, which is equivalent to zero numerically and hence
hard to resolve. A more transparent way is to plot the matrix norm
$||T_{00}(\omega + i0^{+}, g^{-1})|| = \sqrt{\sum_a|\lambda_a(\omega +
  i0^{+}, g^{-1})|^2}$ over the $(\omega,g^{-1})$ parameter space, where
$\lambda_a(z,g^{-1})$ are (complex) eigenvalues of $T_{00}(z,g^{-1})$.
If $||T_{00}(\omega + i0^{+}, g^{-1})||$ is large, then the DOS is in
general enhanced, and one obtains a resonance. This is shown for the
graphene example in Fig.~\ref{gra-tnorm}.

To facilitate analytical treatment, we will henceforth
use the $\ttr^{-1}$ approach and make no distinctions among resonance,
anti-resonance, and bound state.  All of them will simply be referred
to as ``resonance". As needed we will also exhibit $||T(\omega + i0^{+})||$
plots, where anti-resonances are suppressed, as a check.

Consider next the existence of a resonance at an
arbitrary point in $(\omega,g^{-1})$ space, \emph{i.e.} the
condition for at least one eigenvalue of $\ttr^{-1}(\omega, g^{-1})$
to be zero. If $g$ is allowed to be complex, then there
are as many solutions of $g$ at a given $\omega$ as the number of
bands: $g^{-1} = u_a(\omega)$, where $\{u_a(\omega)\}$ are the eigenvalues
of $\grz_{00}(\omega)\Lambda$. However, only real $g$ is physical.
Thus existence of resonance demands at least one eigenvalue of
$\grz_{00}\Lambda$ to be real. The required coupling strength is
$g = 1/u_a(\omega)$. An immediate corollary is that impurities with
$[\Lambda, \grz_{00}] = 0$ can induce a resonance at arbitrary energy
for some $g$, because product of commuting Hermitian matrices has real
eigenvalues. Single-band problems fall in this category ($\Lambda =\id$).

\section{Weyl semimetal models}
\label{sec:model}
We now apply the method described above to lattice systems adapted
from the continuum models of BHB \cite{BHB11}. The $\mib k$-space
Hamiltonian in the $\Gamma$-matrix basis is
\begin{equation}
  \label{hk0}
  \hz(\mib k) = \xi(\mib k) \id + \sum_{i=1}^3 d_i(\mib k) \Gamma^i +
  m(\mib k)\Gamma^4 + \eta\GSB
\end{equation}
where the $\mib k$-dependent coefficients are taken to be
\begin{equation}
  \notag \xi(\mib k) = -2t\sum_{i=1}^3 \cos k_i - \varepsilon_0\ , \ 
  \notag d_i(\mib k) = -2t_1\sin k_i\ ,\ 
  m(\mib k) = -4t\pr\sum_{i=1}^3 (1-\cos k_i) - \lambda\ ,
\end{equation}
and $\eta$ is $\mib k$-independent for simplicity. $\GSB$ is a
$\Gamma$ matrix which breaks time-reversal ($\CT$\,) and/or inversion
($\CI$\,) symmetry to be defined later. The following $\Gamma$ matrix
convention is used:
\begin{equation}
  \label{gamma-lattice}
  \Gamma^i = \tau^x \otimes \sigma^i\ (i = 1,2,3)\ ,\ 
  \Gamma^4 = \tau^z \otimes \id\ , \ \Gamma^5 = -\tau^y \otimes \id\ ,  \ 
  \Gamma^{\mu\nu} = i[\Gamma^{\mu}, \Gamma^{\nu}]/2\ ,
\end{equation}
where $\tau^i$ and $\sigma^i$ are two sets of Pauli matrices acting on
the orbital and spin degrees of freedom, respectively. In this model,
$t$ and $t\pr$ are intra-orbital hoppings, while $t_1$ is the
(spin-mixing) hopping between different orbitals. Different
conventions of $\Gamma$ matrices may have different physical
interpretations: for example, in the convention above, $\Gamma^{12} =
- \id \otimes \sigma^z$ represent a magnetic field in the $z$
direction, but in other conventions it may also have orbital effects.
Results obtained below are independent of the convention used.

\begin{table} 
\centering
\caption{\label{tab:sym} Symmetry of $\Gamma$ matrices. $\CT$ and $\CI$ stand for time-reversal and inversion, respectively.
A plus sign indicates that the associated $\Gamma$ matrices commute with the corresponding symmetry operation, and a minus sign indicates
anticommutation.}
\begin{equation}
\notag
\begin{array}{@{} c c c }
\br
\text{matrices} & \CT & \CI \\
\mr
\id, \Gamma^4 & + & +\\
\Gamma^1, \Gamma^2, \Gamma^3, \Gamma^5 & - & -\\
\Gamma^{12}, \Gamma^{13}, \Gamma^{23}, \Gamma^{15}, \Gamma^{25}, \Gamma^{35} & - & +\\
\Gamma^{14}, \Gamma^{24}, \Gamma^{34}, \Gamma^{45} & + & -\\
\br
\end{array}
\end{equation}
\end{table}

BHB showed how the emergence of stable point or line nodes in the
spectrum of $\hz$ beyond a critical perturbation strength depends on
the symmetry of the perturbation \cite{BHB11}. Define time reversal as
$\CT = \textsf{KR}$ where $\textsf{K}$ is complex conjugation and
$\textsf{R} = \id \otimes i\sigma^y$, and inversion as $\CI =
\Gamma^4$. Symmetry properties of all $\Gamma$ matrices can be found
in Table \ref{tab:sym}. If $\eta = 0$, then the model is both $\CT$
and $\CI$\,-symmetric. In this case, fine-tuning $\lambda = 0$ creates
a Dirac node at $\mib k = 0$ where the four bands converge to the
energy $E = -6t - \varepsilon_0$. While the Dirac point is gapped out
by nonzero $\lambda$, and is therefore unstable in the $\CT$ and
$\CI$\,-symmetric system, it splits into two Weyl nodes if either
$\CT$ or $\CI$ is broken by $\GSB$ \cite{BHB11,Murakami07b}. In this
case the nodal structure survives in a range of parameters and
constitutes a stable nodal phase, \emph{i.e.} the Weyl semimetal (WS)
phase.

In BHB's language, $\hz$ with $\eta = 0$ is the unperturbed
Hamiltonian, and the $\eta\GSB$ term is the symmetry-breaking
perturbation. Although the Weyl nodes and hence the semimetal phase
are stable under such \emph{homogeneous} bulk perturbations, the characteristic
suppression of DOS at the nodal energy in the WS phase, may be
destroyed by another type of perturbation -- namely localized impurity
potentials -- if resonances can be induced at the nodal energy.
Hereafter, we shall refer to the full $\hz$ of Eq. \ref{hk0}, including the
homogeneous term $\eta\GSB$, as the unperturbed Hamiltonian, and
regard the local impurity potential as the perturbation.

The model of Eq. \ref{hk0} is lattice-based, which entails a specific
cutoff structure, and hence not generic like the BHB Hamiltonian
$H_{\BHB}= \sum_{i=1}^3k_i \Gamma^i + m\Gamma^4 + \eta\GSB$. It will
be instructive to first look at the effect of scattering from the more
universal low energy states living in the vicinity of Weyl nodes,
which we shall analyze in Sec.~\ref{sec:cont} for the BHB model.
However, as we shall see there, the very act of taking a momentum
cutoff will leave out the possibility of a stable Weyl node. This is
not surprising because spatially localized impurities are homogeneous
in the momentum space and inevitably scatter high momentum states.
Thus a lattice treatment is necessary. For the lattice theory, we will
first discuss the impurity effect in the $\CT$ and $\CI$\,-symmetric
system ($\eta = 0$). While it does not yield a WS phase, it is simple
enough to be used as a demonstration of the general framework outlined
in Sec.~\ref{sec:method}. Then we will move on to unperturbed systems
with $\CT$ and/or $\CI$ broken ($\eta \neq 0$) where a WS phase does
exist. Since the Hermitian part of the local unperturbed Green's
function, $\grz_{00}$, is of central importance to the impurity
classification, we will first classify the \emph{unperturbed} system
according to the type of $\Gamma$ matrices appearing in $\grz_{00}$,
and then for each of them, classify the impurities according to their
commutation with $\grz_{00}$.

\section{Impurity scattering in the low energy theory}
\label{sec:cont}
In this section we focus on the low energy theory described by the
following BHB Hamiltonian,
\begin{gather}
  H_{\BHB} = \sum_{i=1}^3 k_i \Gamma^i + m \Gamma^4 + \eta \GSB\ .
\end{gather}
Consider for example $\GSB = \Gamma^{21}$ and $\Gamma^{35}$. The
eigenvalues of $H_{\BHB}$ are $\pm E_s$ where
\begin{gather}
  E_s =
  \begin{cases}
    \sqrt{k_x^2 + k_y^2 + \bigl( \eta + s\sqrt{k_z^2 + m^2}\bigr)^2} \quad , \quad \GSB = \Gamma^{21} \\
    \sqrt{k_z^2 + \bigl( \eta + s \sqrt{k_x^2 + k_y^2 + m^2}\bigr)^2} \quad , \quad \GSB = \Gamma^{35}
  \end{cases}\quad , \quad s = \pm 1\ .
\end{gather}
In both cases the scale of nodal momentum is $\Delta = \sqrt{\eta^2 - m^2}$.
For $\eta > m > 0$, $\GSB = \Gamma^{21}$ generates two point nodes at
$\vec k = (0, 0, \pm\Delta)$, whereas $\GSB = \Gamma^{35}$ generates a
nodal line at $\vec k = (\Delta \cos \phi, \Delta \sin\phi, 0)$. In
the vicinity of the Weyl nodes, the $s = -1$ bands have linear
dispersion, whereas the $s = +1$ pair is gapped, $E_+ = 2\eta +
{\cal O}(q^2/\eta^2)$.

If we are looking for low energy resonances, $\omega \sim 0$, then
since the Green's function is weighted by $1/(\omega - E)$, it is
reasonable to focus on the scattering of low energy states by (i)
projecting onto the $s = -1$ bands, and (ii) adopting a momentum
cutoff such that only momenta within a distance $Q$ from the nodes are
considered -- a sphere around point nodes and a tube around nodal line
-- with $\omega \ll Q \ll \eta$. In this approximation the local
Green's function for both $\GSB$s have the following form,
\begin{gather}
  \gz_{00}(\omega) = a(\omega)\left( \id - \frac{m}{\eta}\, \Gamma^4\, \GSB\right)\ ,
\end{gather}
with
\begin{gather}
  a(\omega) =
  \begin{cases}
    \dfrac{Q\omega}{4\pi^2} \dfrac{\eta}{m} \log \dfrac{\eta-m}{\eta+m} - \dfrac{i\eta}{4\pi\Delta}\omega^2 \quad & (\GSB = \Gamma^{21})\\ \\
    \dfrac{\Delta}{2}\, R(m/\eta) \,\omega - \dfrac{i \eta}{8} |\omega| & (\GSB = \Gamma^{35})\ ,
  \end{cases}
\end{gather}
see \ref{sec:app-cont} for derivation and eq.~\ref{rint} for the
expression of $R(m/\eta)$. The hermitian $\grz_{00}(\omega)$
is obtained by taking the real part of $a(\omega)$.

Let us now analyze the resonance condition. The $\ttr^{-1}$ matrix is
\begin{gather}
  \ttr_{00}^{-1}(\omega) = g^{-1}\Lambda + a_r(\omega)\, \frac{m}{\eta}\,
  \Gamma^4\, \GSB - a_r(\omega) \,\id
\end{gather}
where $a_r$ is the real part of $a$. The impurity $\Lambda$ either
commutes or anticommutes with $\Gamma^4\,\GSB$ since the latter is
itself one of the sixteen $\Gamma$ matrices. If they commute, then
$\det\, \ttr_{00}^{-1} = 0$ yields real solutions for $g^{-1}$,
\emph{i.e.} resonance could be induced and $\omega \sim 0$ is
unstable. If, on the other hand, $\Lambda$ and $\Gamma^4\,\GSB$
anticommute, then $\det \,\ttr_{00}^{-1} = 0$ gives
\begin{gather}
  g^{-1} = \pm \big|a_r(\omega)\big|\sqrt{1 - \frac{m^2}{\eta^2}}\ .
\end{gather}
Now, by $m^2 + \Delta^2 = \eta^2$, one obtains $g^{-1} = \pm \big|a_r(\omega)\big|
\Delta / \eta$, which is still real, \emph{i.e.}, $\omega \sim 0$ is
unstable for anticommuting $\Lambda$. Thus in the cutoff scheme
adopted here, $\omega \sim 0$ is unstable regardless of the type of
impurity. Note however that in the case of anticommuting ones, the
impurity strength $g$ always comes in $\pm$ pairs, whereas in the
commuting case it might not.

An essential difference between the commuting and anticommuting
impurities is that in the latter case, the solution for $g^{-1}$
contains a square root. When higher momentum states are considered,
the argument of the square root may become negative, and stabilize the
nodal energy. 

To see this, note that without the low energy restriction, the local
Green's function of the BHB Hamiltonian with $\GSB = \Gamma^{21}$ and
$\Gamma^{35}$ has the form (see eq.~\ref{coef:gmunu} in
\ref{sec:app-g})
\begin{gather}
  \gz_{00}(\omega) = a(\omega)\, \id + b_1(\omega)\, \Gamma^4 +
  b_2(\omega)\, \GSB + b_3(\omega)\, \Gamma^4\, \GSB
\end{gather}
where the coefficients are
\begin{gather}
  \label{g12-a}
  a(\omega) =  \frac{\omega}{2} \left< \frac{1}{\omega^2 - E_+^2} + \frac{1}{\omega^2 - E_-^2}\right>\ , \\
  \label{g12-b1}
  b_1(\omega) = \frac{m}{2} \left< \frac{1}{\omega^2 - E_+^2} + \frac{1}{\omega^2 - E_-^2} + \frac{4\eta^2}{(\omega^2 - E_+^2)(\omega^2 - E_-^2)} \right> \ ,\\
  \label{g12-b2}
  b_2(\omega) =  \frac{\eta}{2} \left< \frac{1}{\omega^2 - E_+^2} + \frac{1}{\omega^2 - E_-^2} + \frac{4 d_{\perp}^2}{(\omega^2 - E_+^2)(\omega^2 - E_-^2)} \right>\ ,\\
  \label{g12-b3}
  b_3(\omega) = 2\eta \omega m \left< \frac{1}{(\omega^2 - E_+^2)(\omega^2 - E_-^2)} \right>\ ,
\end{gather}
in which $\langle \cdots \rangle$ denotes the $\vec k$-space integral
(for continuum) or sum (for lattice), and
\begin{gather}
  d_{\perp}^2 =
  \begin{cases}
    k_z^2 + m^2 & (\GSB = \Gamma^{21})\\
    k_x^2 + k_y^2 + m^2 & (\GSB = \Gamma^{35})
  \end{cases}\ .
\end{gather}
In the low energy approximation, $b_1$ and $b_2$ vanish due to the
momentum cutoff. This is because at the Weyl nodes, $E_+^2 =
d_{\perp}^2 = 4\eta^2$, thus in the low energy approximation, the
second and third terms inside $\langle \cdots \rangle$ of both $b_1$
and $b_2$ cancel, yielding
\begin{gather}
  b_1(\omega) \simeq \frac{m}{2(\omega^2 - E_+^2)}
  \frac{V_{\textsf{cutoff}}}{V_{\textsf{BZ}}}\quad , \quad b_2(\omega) \simeq \frac{\eta}{2(\omega^2 - E_+^2)} \frac{V_{\textsf{cutoff}}}{V_{\textsf{BZ}}}\ ,
\end{gather}
where $V_{\textsf{cutoff}}/V_{\textsf{BZ}}$ is the ratio between the
volume within the momentum cutoff and that of the first Brillouin
zone. This volume ratio comes from the evaluation of $\langle 1
\rangle$, and is of the order $(Q/2\pi)^{\textsf{codim}} \sim
(Q/\eta)^{\textsf{codim}} \rightarrow 0$ in the low energy
approximation, with $\textsf{codim}$ being the codimension of the Weyl
node, which is $3$ for a point node and $2$ for a nodal line. When
higher momentum states are included, $b_1$ and $b_2$ will no longer be
suppressed, and will change the argument under the square root,
possibly making it negative and stabilizing the nodal energy. A more
careful analysis necessitates a lattice treatment, which is what we
shall do in the rest of the paper.

\section{Classification of impurity potentials}
\label{sec:imp}

\subsection{$\CT$ and $\CI$\,-symmetric $\hz(\mib k)$ }
\label{sec:ti-sym}
We now return to the lattice model of Sec.~\ref{sec:model}. To
illustrate the resonance criteria of Sec.~\ref{sec:method}, we first
consider the case with $\eta = 0$. Inverting Eq.~\ref{hk0} yields the
unperturbed $\mib k$-space Green's function,
\begin{equation}
  \gz(\omega, k) = \frac{\left[\omega - \xi(\mib k)\right] \id +
    \sum_{i=1}^3 d_i(\mib k)\Gamma^i + m(\mib k)\Gamma^4}{\left[\omega
      - \xi(\mib k)\right]^2 - \sum_{i=1}^3|d_i(\mib k)|^2 - m^2(\mib k)}
\end{equation}
The $\Gamma^i$ ($i = 1,2,3$) terms will vanish after summation over
$\mib k$ due to the oddness of $d_i(\mib k)$, as required by inversion
symmetry. The local Green's function is thus
\begin{equation}
  \label{gz00-h0}
  \gz_{00}(\omega) = \frac{1}{N} \sum_{\mib k} \gz(\omega, \mib k) =
  a(\omega) \id + b(\omega)\Gamma^4\ ,
\end{equation}
where $N$ is the number of $\mib k$ points, 
\begin{equation}
  \label{gz00-h0-ab}
a(\omega) = {1\over N}\sum_{\mib k} \frac{\omega - \xi(\mib k)}{D(\omega, \mib k)} \qquad,\qquad
b(\omega) = {1\over N}\sum_{\mib k} \frac{m(\mib k)}{D(\omega, \mib k)}\quad,
\end{equation}
and
\begin{equation}
  \label{gz00-h0-d}
D(\omega,\mib k) = \left[\omega - \xi(\mib k)\right]^2 - \sum_{i=1}^3 d_i^2(\mib k) - m^2(\mib k)\ .
\end{equation}

As discussed in Sec.~\ref{sec:method}, the existence of a resonance
depends on whether or not the eigenvalues of $\ttr_{00}^{-1}$,
\emph{i.e.} the Hermitian part of the inverse local $T$-matrix, can be
zero. Since the only $\Gamma$ matrix in the $\grz_{00}$ decomposition is $\Gamma^4 =
\CI$, there are only two classes of impurities according to their
inversion property:

\subsubsection{Inversion-even impurity}

\begin{figure}
  \includegraphics[width=0.5\textwidth,]{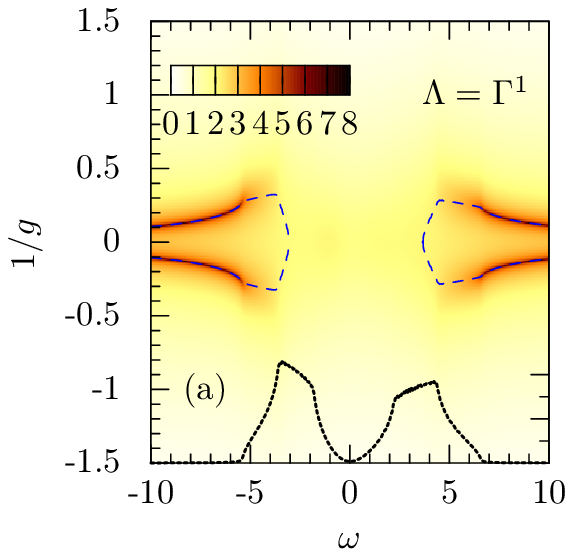}
  \includegraphics[width=0.41\textwidth,]{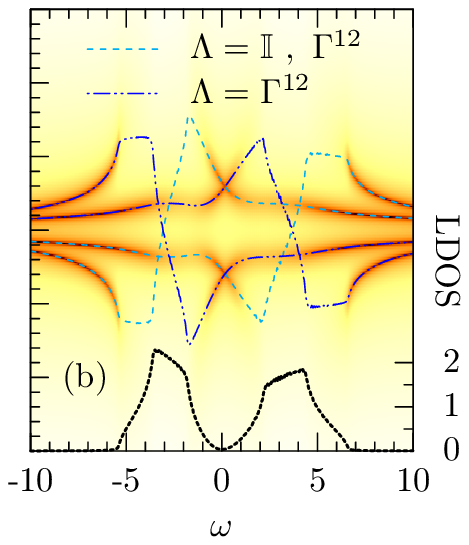}
    \caption{(Color online) Color map: $\log || T_{00}(\omega +
      i\epsilon)||$ for $\CT$ and $\CI$\,-symmetric case
      ($\eta = 0$). Darker color corresponds to stronger impurity
      effect. Unperturbed DOS is shown at the bottom. Colored lines
      interpolating the dark curves are obtained by replacing
      $\gz_{00}$ with $\grz_{00}$ and computing the zeros of
      $g^{-1}\Lambda - \grz_{00}(\omega)$ in the $g-\omega$ plane (see
      text). For $\Lambda=\Gamma^1$ (a), the Dirac node is stable.
      Panel (b) shows results for $\Lambda = \Gamma^{12}$ (all curves)
      and for $\Lambda=\id$ (blue dashed curves only); the Dirac node
      is unstable. Parameters used are $t = 0.05, t_1 = -0.5,
      t^{\prime} = -0.25, \lambda = 0, \varepsilon_0 = -0.3$, and
      lattice size $N_x = N_y = N_z = 50$. Spectral broadening
      $\epsilon$ is set to $0.05$.}
  \label{tnorm-h0}
\end{figure}

In this class we have $\Lambda = \id, \Gamma^4,$ or $\Gamma^{\mu\nu}$
($\mu,\nu = 1,2,3,5$). Since they all commute with $\grz_{00}$,
a resonance can be induced at arbitrary energy, {\it i.e.\/} a solution of $\det\, \ttr_{00}^{-1} = 0$
exists for real $g$. To illustrate, we solve for $g(\omega)$, the value of $g$ which produces a resonance
at energy $\omega$, for all three cases:
\begin{itemize}
\item[(1)] $\Lambda = \id$\,: This is a scalar impurity, and
\begin{equation}
  \ttr_{00}^{-1}(\omega) = g^{-1} - \grz_{00}(\omega) = \left[g^{-1} -
  a(\omega)\right]\id - b(\omega)\Gamma^4\ .
\end{equation}
The \emph{principal values} of $a$ and $b$ \emph{are implicitly
  taken}. Setting the LHS to zero yields
\begin{equation}
  g^{-1}(\omega) = a(\omega) \pm b(\omega)\ .
\end{equation}
These are shown as the light blue dashed lines in Fig.~\ref{tnorm-h0}(b).

\item[(2)] $\Lambda = \Gamma^4$\,: This impurity flips the sign of the inversion-odd component, yielding
\begin{equation}
  \ttr_{00}^{-1}(\omega) = - a(\omega)\,\id + \left[g^{-1} -
    b(\omega)\right]\Gamma^4\ .
\end{equation}
The resonance condition is thus
\begin{equation}
  g^{-1}(\omega) = b(\omega) \pm a(\omega)\ .
\end{equation}

\item[(3)]$\Lambda = \Gamma^{\mu\nu}$ ($\mu,\nu = 1,2,3,5$)\,: This includes
for example the magnetic impurities, $\Gamma^{12} = - \id\otimes
\sigma^z$, etc, and
\begin{equation}
  \ttr_{00}^{-1}(\omega) = g^{-1}\Lambda - a(\omega) \,\id - b(\omega)\,\Gamma^4\ .
\end{equation}
The eigenvalues of $\ttr_{00}^{-1}$ are obtained by replacing
$\Lambda$ and $\Gamma^4$ on the RHS each with uncorrelated $\pm 1$
(since they can be simultaneously diagonalized). Setting these
eigenvalues to zero yields
\begin{equation}
  g^{-1}(\omega) = \pm a(\omega) \pm b(\omega)\ .
\end{equation}
These are shown as dash and dash-dot lines in Fig.~\ref{tnorm-h0}(b).
\end{itemize}

\subsubsection{Inversion-odd impurity}
In this class we have $\Lambda = \Gamma^\mu$ or $\Gamma^{4\mu}$ with $\mu =
1,2,3,5$. The inverse $T$ matrix is
\begin{equation}
  \ttr_{00}^{-1}(\omega) = g^{-1}\Lambda - a(\omega)\id - b(\omega)\Gamma^4\ .
\end{equation}
Rearranging and squaring, one gets $(\ttr_{00}^{-1} + a\id)^2 = g^{-2}
+ b^2$. The cross term of $g^{-1}$ and $b$ vanishes because
$\{\Lambda, \Gamma^4\} = 0$ . Setting $\ttr_{00}^{-1} = 0$ then yields
\begin{equation}
  g^{-1}(\omega) = \pm \sqrt{a^2(\omega) - b^2(\omega)}\ .
\end{equation}
The results are plotted as dashed lines in Fig.~\ref{tnorm-h0} (a). Thus
inversion-odd impurities cannot induce any resonance in the range of
energy $\omega$ for which
\begin{equation}
  \label{h0-stable}
  |a(\omega)| < |b(\omega)|\ .
\end{equation}

\subsubsection{Band center approximation (BCA)}
\label{sec:BCA}

To understand the general trend of the resonance solutions $g(\omega)$
and get a sense of the stability region, it is useful to obtain an
approximation for the expansion coefficients $a(\omega)$ and
$b(\omega)$. To this end we introduce the \emph{band center
  approximation} (BCA): A generic $\mib k$-space Hamiltonian $H(\mib
k)$ can be written as $H(\mib k) = H_{00} + \delta H_{\mib k}$ where
$H_{00} = \langle H(\mib k)\rangle$ is the local Hamiltonian and
$\langle \cdots \rangle$ denotes $\mib k$-space averaging. The
eigenvalues of $H_{00}$ can be thought of as some sort of average
energy of the bands of $H(\mib k)$ (band centers). Let
\begin{equation}
  \bar G(\omega) = (\omega - H_{00})^{-1}\ ,
\end{equation}
then the local Green's function is
\begin{align}
  \label{BCA}
  \notag
  G_{00} &= \langle (\omega - H_{00} - \delta H_{\mib k})^{-1}\rangle \\
  \notag
  &= \bar G + \bar G \langle \delta H_{\mib k}\rangle \bar G + \bar G\,
  \langle \delta H_{\mib k} \ \bar G\ \delta H_{\mib k}\rangle \,\bar G
  + \cdots\\
  &= \bar G\,\Big\{ 1 + {\cal O}\!\left((\delta H_{\mib k}\, \bar G)^2\right)\Big\}
\end{align}
where we have used $\langle \delta H_{\mib k}\rangle = 0$. The BCA
amounts to replacing the local Green's function, $G_{00}$, with the
Green's function of the local Hamiltonian, $\bar G$. 

Eq.~\ref{BCA} is an expansion in powers of $\delta H_{\mib k}/(\omega
- H_{00})$, where the numerator is roughly the band width,
and the denominator is the distance from $\omega$ to the band centers.
The BCA works well if the distance of $\omega$ from some band center, say that of
band $A$, is greater than $A$'s bandwidth.  Note that such an $\omega$,
although outside band $A$, may well be inside another band, say band
$B$. From the BCA point of view, a resonance at some $\omega$ inside
band $B$ is actually a consequence of the coherent superposition of
states mainly in some other band ($A$). The multiple-band scenario is
to the benefit of the BCA.

\begin{table}
\caption{\label{tab:ibreak} Impurity classification for $\CI$ breaking Weyl material. The symmetry breaking term in $\hz(\mib k)$ is $\GSB = \Gamma^{\mu 4}$ ($\CT$ even) or $\Gamma^{\mu}$ ($\CT$ odd) where $\mu \in \{1,2,3,5\}$. The second column indicates commutation ($+$) or anticommutation ($-$) of the impurity matrix $\Lambda$ with $\Gamma^4$ and $\GSB$, respectively. Elements in each class are enumerated in the first column: if the cell has two sub-cells, the left one corresponds to $\GSB = \Gamma^{\mu 4}$ and right one $\GSB = \Gamma^{\mu}$; otherwise the enumeration is identical for both $\GSB$. The value of $g$ at which $\ttr_{00}^{-1}$ has a zero eigenvalue is listed in the third column, and the condition for it to be real (the resonance condition) is shown in the fourth column. The fifth column shows the resonance condition as given by the band center approximation, which are simple expressions in terms of the Hamiltonian parameters.  Note that the values of the Green's function coefficients $a(\omega)$ and $b_i(\omega)$ depend on the choice of $\GSB$ that breaks $\CI$, but the BCA conditions are \emph{independent} of $\GSB$. The stability of Weyl nodes, if they exist, is listed in the last column.}
\begin{equation*}
\begin{array}{ @{} l  l  c   r  c  c  c }
\br
\multicolumn{2}{c}{\Lambda\ (\Gamma^{\mu 4} | \Gamma^{\mu})} & \text{class}  & \multicolumn{1}{c}{g^{-1}} & \text{resonance} & \text{resonance (BCA)} & \text{node stability} \\
\mr
\multicolumn{2}{c}{\Gamma^4} & \multirow{2}{*}{$(+,-)$} & b_1 \pm \sqrt{a^2 - b_2^2} & \multirow{2}{*}{$|a| > |b_2|$} & \multirow{2}{*}{$|\omega - \alpha| > |\eta|$} & \multirow{6}{*}{stable} \\
\multicolumn{2}{c}{\Gamma^{\mu \bar{\mu}}} & & \pm b_1 \pm \sqrt{a^2 - b_2^2} & & & \\
\cline{1-3}\cline{4-6}
\Gamma^{\mu 4}\;\; & \Gamma^{\mu} & \multirow{2}{*}{$(-,+)$} & b_2 \pm \sqrt{a^2 - b_1^2} & \multirow{2}{*}{$|a| > |b_1|$} & \multirow{2}{*}{$|\omega - \alpha| > |\beta|$} & \\
\Gamma^{\bar{\mu}} & \Gamma^{\bar{\mu} 4}& & \pm b_2 \pm \sqrt{a^2 - b_1^2} & & &\\
\cline{1-6}
\Gamma^{\mu} & \Gamma^{\bar{\mu}}& \multirow{2}{*}{$(-,-)$} & \multirow{2}{*}{$\pm \sqrt{a^2 - b_1^2 - b_2^2}$} & \multirow{2}{*}{$a^2 > b_1^2 + b_2^2$} & \multirow{2}{*}{$(\omega - \alpha)^2 > \beta^2 + \eta^2$} & \\
\Gamma^{\bar{\mu} 4} & \Gamma^{\mu 4}& & & & &\\
\hline
\multicolumn{2}{c}{\id} & \multirow{2}{*}{$(+,+)$} & a \pm \sqrt{b_1^2 + b_2^2} & \multirow{2}{*}{any $\omega$} & \multirow{2}{*}{any $\omega$} & \multirow{2}{*}{unstable} \\
\multicolumn{2}{c}{\Gamma^{\bar{\mu} \bar{\nu}}} & & \pm a \pm \sqrt{b_1^2 + b_2^2} & & &\\
\br
\end{array}
\end{equation*}
\begin{equation*}
\mu \in \{1,2,3,5\}\ (\text{fixed in }\hz)\ ,\ \bar\mu, \bar \nu
\in \{1,2,3,5\}
\setminus \{\mu\}\ .\ 
\alpha = - \varepsilon_0\ , \ \beta = -12 t\pr - \lambda\ .
\end{equation*}
\end{table}
Applying BCA to the $\eta = 0$ model here, one finds from
Eq.~\ref{hk0} that $H_{00} = \alpha \id + \beta \Gamma^4$ where
$\alpha = -\varepsilon_0$ and $\beta = - 12t\pr - \lambda$, hence
\begin{equation}
  \bar G = \bar{a}\id + \bar{b}\Gamma^4\qquad,\qquad
  \bar{a} = \frac{\omega - \alpha}{(\omega - \alpha)^2 - \beta^2}\qquad,\qquad
  \bar{b} = \frac{\beta}{(\omega - \alpha)^2 - \beta^2}\ .
\end{equation}
$\bar G$ and $G_{00}$ have the same form of decomposition. This will
prove useful in the more complicated situations where $\GSB$ is
present.

For the scalar potential $\Lambda = \id$, the BCA resonance solution
$g^{-1} \simeq \bar a(\omega) \pm \bar b(\omega) = \left[\omega -
  (\alpha \pm \beta)\right]^{-1}$ resembles two hyperbolae centered around the
band centers $\alpha \pm \beta$. They can be identified qualitatively
from the light blue dashed curves in Fig.~\ref{tnorm-h0}(b), although
numerically the two branches of each hyperbola, instead of being
divergent, are connected around their respective band centers due to
higher order effects in Eq.~\ref{BCA}. For $\CI$ odd impurities, such as
the magnetic impurity $\Lambda = \Gamma^{12}$, the stability condition
Eq.~\ref{h0-stable} implies $\alpha - |\beta| < \omega < \alpha +
|\beta|$, \emph{i.e.}, stable energy $\omega$ is bounded by the two
band centers, as can be seen from Fig.~\ref{tnorm-h0} (a). This region
in particular includes the (fine-tuned) Dirac point or the central
gap. We thus conclude that the Dirac node is generically stable for
$\CI$ odd impurities and unstable for $\CI$ even impurities.

\subsection{$\hz(\mib k)$ with $\CI$ breaking $\GSB$}
\label{sec:ibreak}
\begin{figure*}
  \centering
  \begin{subfigure}[t]{0.32\textwidth}
    \includegraphics[width=\textwidth]{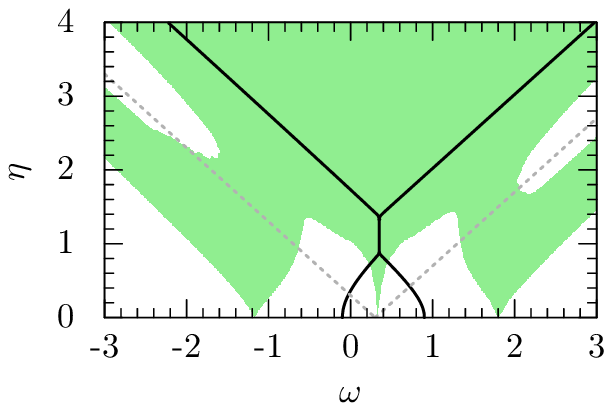}    
    \caption{class $(+,-)$}
  \end{subfigure}
  \begin{subfigure}[t]{0.32\textwidth}
    \includegraphics[width=\textwidth]{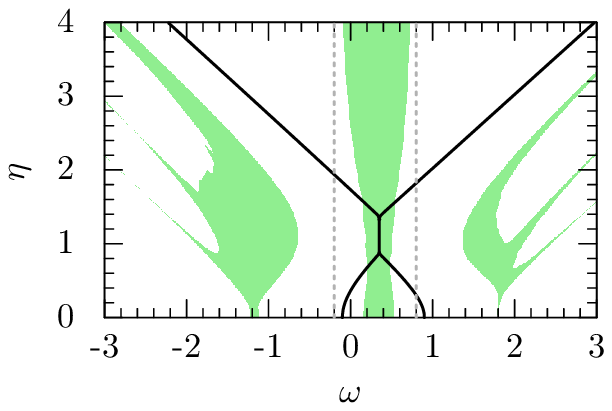}    
    \caption{class $(-,+)$}
  \end{subfigure}
  \begin{subfigure}[t]{0.32\textwidth}
    \includegraphics[width=\textwidth]{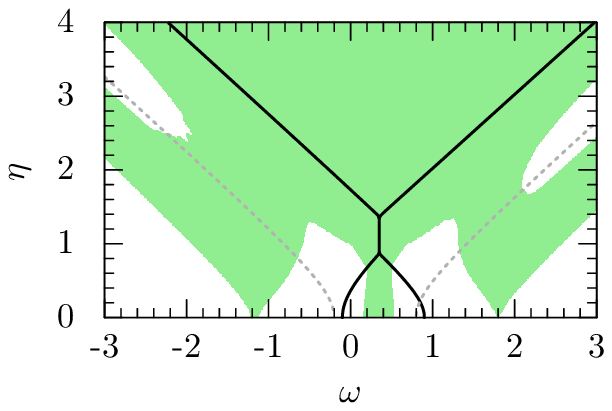}
    \caption{class $(-,-)$}
  \end{subfigure}
  \caption{(Color online) Stability of Weyl semimetal with
    $\CI$ breaking $\GSB = \Gamma^{14}$ (bulk perturbation
    with both magnetic and orbital effects). The vertical axis $\eta$
    is the strength of the $\CI$ breaking term. (a) stable zone for
    impurity classes $(+,-)$. (b) stable zone for impurity classes
    $(-,+)$. (c) stable zone for impurity class $(-,-)$. See Table
    \ref{tab:ibreak} for impurity classification. Note how (c)
    resembles the union of (a) and (b), as could be predicted from
    Table \ref{tab:ibreak}. Solid black lines mark the two band edges
    bounding the central gap. They touch from around $\eta = 0.88$ to
    $1.35$, corresponding to the Weyl semimetal phase. Dotted gray
    lines are the stable zone boundaries given by the band center
    approximation. It qualitatively agrees with the shape of the
    colored region near the central gap. The deviation is mainly deep
    in the bands (side wings in the colored region) where higher order
    terms in Eq.~\ref{BCA} become important. Parameters used are $t =
    0.05, t_1 = -0.5, t^{\prime} = -0.25, \lambda = 3.5, \varepsilon_0
    = -0.3$, on a lattice of $N_x = N_y = N_z = 50$. Spectral
    broadening is $\epsilon = 0.05$. }
  \label{ibreak-phase}
\end{figure*}

We now consider the case where $\GSB$ breaks inversion. In the
BHB scheme, this can be realized for example by applying a voltage
bias across each TI layer, breaking the inversion symmetry between the
two TI surfaces. According to Table \ref{tab:sym}, $\GSB =
\Gamma^{\mu 4}$ ($\CT$ even) or $\Gamma^{\mu}$ ($\CT$ odd) where
$\mu = 1,2,3,5$. The local Green's function has the decomposition,
\begin{equation}
  \label{gz-ibreak}
  \gz_{00}(\omega) = a(\omega)\,\id + b_1(\omega)\,\Gamma^4 +
  b_2(\omega)\,\GSB\ .
\end{equation}
While this decomposition can be obtained analytically (see
Eqs.~\ref{coef:gmu4}-\ref{coef:gmu4-end} and
Eqs.~\ref{coef:gmu}-\ref{coef:gmu-end}), its structure is easier to
understand from the BCA: symmetry consideration demands that the local
Hamiltonian $\hz_{00} \equiv \frac{1}{N}\sum_{\mib k}\hz(\mib k) =
\alpha\, \id + \beta\, \Gamma^4 + \eta\, \GSB$ where the first two
terms are the only possibilities to conserve both $\CT$ and $\CI$,
see Table \ref{tab:sym}. Its inverse can potentially have four terms,
$\id$, $\Gamma^4$, $\GSB$, and $\Gamma^4\,\GSB$. Since
$\GSB$ anticommutes with $\Gamma^4$, their cross term must
vanish, yielding the form in Eq.~\ref{gz-ibreak}. We will come back to
BCA later.

The resonance condition can now be solved for different impurities.
As an example, consider $\GSB = \Gamma^{\mu 4}$ and $\Lambda =
\Gamma^{\mu \bar \mu} = \Lambda^{-1}$ where $\bar{\mu} \in \{1,2,3,5\}
\setminus \{\mu\}$. This type commutes with $\Gamma^4$ but
anticommutes with $\GSB$, and includes the purely magnetic
impurities $\Gamma^{12}, \Gamma^{23}$ and $\Gamma^{13}$. The Hermitian
part of the inverse $T_{00}$ matrix is
\begin{equation}
  \ttr_{00}^{-1}(\omega) = g^{-1} \Gamma^{\mu \bar{\mu}} - a(\omega) \id - b_1(\omega) \Gamma^4 - b_2(\omega) \Gamma^{\mu 4}\ .
\end{equation}
Using the anticommutation $\{g^{-1}\Gamma^{\mu\bar{\mu}} -
b_1\Gamma^4, \Gamma^{\mu 4}\} = 0$, the above can be rearranged into
$\left(\ttr_{00}^{-1} + a\right)^2 - b_2^2 = \left(g^{-1} \Gamma^{\mu
    \bar{\mu}} - b_1\Gamma^4\right)^2$. Setting $\ttr_{00}^{-1} = 0$,
both sides can be simultaneously diagonalized, and the eigenvalues of
the RHS are $(g^{-1} \pm b_1)^2$. The condition for vanishing
$\ttr_{00}^{-1}$ is thus
\begin{equation}
  g^{-1} = \pm b_1(\omega) \pm \sqrt{a(\omega)^2 - b_2(\omega)^2}\ .
\end{equation}
Resonance then requires $g^{-1}$ to be real, \emph{viz.} $|a(\omega)|
> |b_2(\omega)|$. The occurrence of a possibly negative term under the
square root stems from the anticommutation of $\GSB$ with
$\Lambda$, \emph{i.e.} the interplay between the bulk symmetry
breaking field and the impurity.

Similar analysis can be carried out when $\Lambda$ is any of the
sixteen $\Gamma$ matrices. The results are summerized in the third
column of Table \ref{tab:ibreak}. The sixteen $\Gamma$-matrix impurity
candidates can be classified into four classes labeled by their
commutation with $\Gamma^4$ and $\Gamma^{\mu 4}$: $(+,-)$ denotes
$\Lambda$ commuting with $\Gamma^4$ and anticommuting with
$\Gamma^{4\mu}$, and similarly for $(+,+)$, $(-,+)$ and $(-,-)$.
Impurities belonging to the same class have the same resonance
condition. A nontrivial solution arises if there is at least one
anticommutation, giving rise to a possibly negative term under the square
root, and the protection of DOS suppression at Weyl nodes. The
unperturbed $\hz$ is parameterized by the symmetry-breaking strength
$\eta$, and one can ask how it affects the system's ability to induce
resonance at energy $\omega$. The $(\omega,\eta)$ space is thus
divided into two phases according to the existence of resonance. These
are shown in Fig.~\ref{ibreak-phase}, where the stable phases (no
resonance) are colored.

The shape of the phase boundaries can be qualitatively understood in
terms of the parameters of the Hamiltonian using BCA: the local
Hamiltonian is $\hz_{00} = \frac{1}{N}\sum_{\mib k} \hz(\mib k) =
\alpha \id + \beta\Gamma^4 + \eta \Gamma^{\mu 4}$ where $\alpha =
-\varepsilon_0\ , \ \beta = -12 t\pr - \lambda$. Its Green's function
is $\bar G(\omega) = \bar a(\omega) \id + \bar b_1(\omega)\Gamma^4 +
\bar b_2(\omega)\Gamma^{\mu 4}$ where $\bar a(\omega) = (\omega -
\alpha)/Q(\omega)$, $\bar{b}_1(\omega) = \beta/Q(\omega)$,
$\bar{b}_2(\omega) = \eta/Q(\omega)$, and $Q(\omega) = (\omega -
\alpha)^2 - \beta^2 - \eta^2$. Note that these coefficients are
\emph{independent} of $\GSB$, thus the stability of the Weyl
nodes can be predicted according to the impurity class, regardless of
$\GSB$. The BCA version of the phase boundaries are shown as
dotted lines in Fig.~\ref{ibreak-phase}. The DOS suppression at the
bulk Weyl nodes is protected for impurities in classes $(+,-)$,
$(-,+)$ and $(-,-)$. The only unstable class is $(+,+)$, due to its
fully-commuting nature with $\grz_{00}$.

\subsection{$\hz(\mib k)$ with $\CI$\,-symmetric $\GSB$}
\label{sec:tbreak}

To split the Dirac node into two Weyl nodes, an $\CI$\,-symmetric
$\GSB$ must break $\CT$. Thus $\GSB = \Gamma^{\mu\nu}$
($\mu \neq \nu \neq 4$) according to Table \ref{tab:sym}. The local
Green's function is
\begin{equation}
  \label{gz-tbreak}
  \gz_{00}(\omega) = a(\omega) \id + b_1(\omega)\Gamma^4 +
  b_2(\omega)\Gamma^{\mu\nu} + b_3(\omega)\Gamma^4\Gamma^{\mu\nu}\ ,
\end{equation}
see Eqs.~\ref{coef:gmunu}-\ref{coef:gmunu-end} for expressions of the
coefficients. The decomposition structure is easier to understand in
terms of BCA: similar to the discussion beneath Eq.~\ref{gz-ibreak},
one has $\hz_{00} = \alpha \id + \beta\Gamma^4 + \eta \GSB$,
thus its inverse has four possible terms, $\id, \Gamma^4, \GSB$
and $\Gamma^4\,\GSB$. Since $\Gamma^4$ and $\GSB$ commute
($\CI$ symmetry), their cross term does not vanish, hence the form of
Eq.~\ref{gz-tbreak}.

\begin{table}
\caption{\label{tab:tbreak} Impurity classification for $\CI$\,-symmetric (hence $\CT$ breaking) Weyl material. The class to which $\Lambda$ belongs are labeled by the three signs of the commutation of $\Lambda$ with $\Gamma^4$, $\Gamma^{\mu\nu}$, and $\Gamma^4\Gamma^{\mu\nu}$ in that order, where $+$ denotes commute and $-$ anticommute. The two indices $\mu, \nu \in \{1,2,3,5\}$ and are fixed by the unperturbed Hamiltonian. The index $p \in \{1,2,3,5\} \setminus \{\mu,\nu\}$. $s$ and $s\pr$ take the values of $\pm 1$. Solution of $g$ yielding $\det\,\ttr^{-1} = 0$ are summerized in the third column, see equations~\ref{res-tbreak-full-com} and \ref{res-tbreak} in text. The resonance conditions for each class can be deduced by requiring $g^{-2}$ to be positive (so that $g$ is real), and are explicitedly spelled out in the fourth column using BCA, which only need to be satisfied for either $s = 1$ or $-1$. Stability of the Weyl nodes are listed in the last column. $\alpha = -\varepsilon_0$ and $\beta = -12t\pr - \lambda$.}
\begin{tabular}{@{} c  c  c  c c}
\br
$ \Lambda$ & class & $g^{-2}$ & resonance (BCA) & stability\\
\mr
$\id$, $ \Gamma^4$, $ \Gamma^{\mu\nu}$, $\Gamma^4\Gamma^{\mu\nu}$ & $(+,+,+)$ &  $(a + s b_1 + s\pr b_2 + ss\pr b_3)^2$ & any $\omega$ & unstable \\
$ \Gamma^{\mu p}$, $ \Gamma^{\nu p}$ & $(+,-,-)$ &  $(a+sb_1)^2 - (b_2 + s b_3)^2$ & $|\omega - \alpha - s\beta | > |\eta|$ & $|\eta| > |\beta|$\\
$ \Gamma^p$, $\Gamma^{4p}$ & $(-,+,-)$ &  $ (a+s b_2)^2 - (b_1 + s b_3)^2$ & $|\omega - \alpha - s\eta | > |\beta|$ & $|\eta| < |\beta|$\\
$\Gamma^{\mu}, \Gamma^{4\mu},\Gamma^{\nu}, \Gamma^{4\nu}$ & $(-,-,+)$ & $ (a+s b_3)^2 - (b_1 + s b_2)^2$ & $|\omega - \alpha| > |\beta + s\eta|$ & stable \\
\br
\end{tabular}
\end{table}

Note that all three $\Gamma$ matrices in Eq.~\ref{gz-tbreak} mutually
commute, and the product of any two is equal to the third. This
implies that the impurity $\Lambda$ either commutes with all of them,
or it commutes with one and anticommutes with the other two (because
the product of any two commutation signs should produce the third). In
the fully commuting case, resonance can always be induced by
impurities of strength
\begin{equation}
  \label{res-tbreak-full-com}
  g^{-1} = s_{\Lambda} \Big[a(\omega) + s_4\, b_1(\omega) + s_{\mu\nu}\, b_2(\omega) + s_4\,s_{\mu\nu}\,b_3(\omega)\Big]
\end{equation}
where $s_4, s_{\mu\nu}$ and $s_{\Lambda}$ are eigenvalues of
$\Gamma^4, \Gamma^{\mu\nu}$ and $\Lambda$ respectively and
independently take the values $\pm 1$.

\begin{figure}
  \centering
  \includegraphics[width=0.6\textwidth]{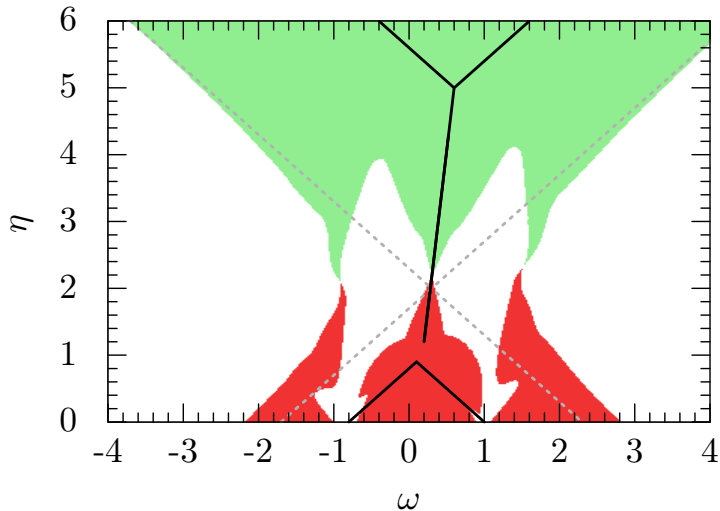}
  \caption{(Color online) Stability of Weyl semimetal with
    $\CI$\,-symmetric and $\CT$ breaking $\GSB = \Gamma^{12}$
    (external magnetic field). The vertical axis $\eta$ is the
    strength of the $\CT$ breaking term. Green (light shade): stable
    zone for impurity classes $(+,-,-)$ and $(-,-,+)$. Red (dark
    shade): stable zone for impurity classes $(-,+,-)$ and $(-,-,+)$.
    See Table \ref{tab:tbreak} for impurity classification. Solid
    black lines mark the two band edges bounding the central gap. They
    touch from around $\eta = 1$ to $5$, corresponding to the Weyl
    semimetal phase. The black lines are broken around $\eta = 1$ due
    to the closing of the indirect gap. Dotted gray lines are the
    stable zone boundaries given by the band center approximation.
    Parameters used are $t = 0.05, t_1 = -0.5, t^{\prime} = -0.25,
    \lambda = 1, \varepsilon_0 = -0.3$, on a lattice of $N_x = N_y =
    N_z = 50$. Spectral broadening is $\epsilon = 0.05$. }
  \label{tbreak-phase}
\end{figure}

For other $\Lambda$, there are two anticommutations. We can relabel
the three $\Gamma$ matrices in Eq.~\ref{gz-tbreak} according to their
commutation with $\Lambda$, and write the inverse $T$ matrix as
\begin{equation}
  \label{t00-tbreak}
  \ttr_{00}^{-1} = - (a \id + b_C\Gamma_C) + (g^{-1}\Lambda -
  b_A\Gamma_A - b_{A\pr}\Gamma_{A\pr})
\end{equation}
where $[\Lambda, \Gamma_C] = \{\Lambda, \Gamma_A\} = \{\Lambda,
\Gamma_{A\pr}\} = 0$ and $\{\Gamma_C, \Gamma_A, \Gamma_{A\pr}\}$ is
some permutation of $\{\Gamma^4, \Gamma^{\mu\nu},
\Gamma^{4}\Gamma^{\mu\nu}\}$. The two parentheses in
Eq.~\ref{t00-tbreak} mutually commute, thus $\ttr_{00}^{-1}$ is
block-diagonal: in the eigen-subspace of $\Gamma_C$ with eigenvalue
$\pm 1$, the matrices $\Gamma_C$, $\Lambda$, $\Gamma_A$ and
$\Gamma_{A\pr}$ reduce to $2\times 2$ blocks, denoted as $\pm \id$,
$\Lambda^{\pm}$, $\Gamma_A^{\pm}$ and $\Gamma_{A\pr}^{\pm}$,
respectively, all of which square to $\id$. Since the projectors onto
the subspaces of $\Gamma_C$ commute with $\Gamma_C, \Gamma_A,
\Gamma_{A\pr}$ and $\Lambda$, the mutual (anti)commutation relations
of the latter four are inherited in both subspaces. Setting
$\ttr_{00}^{-1} = 0$ in eq.~\ref{t00-tbreak} for both blocks then
yields
\begin{equation}
  (a \pm b_C) \id = g^{-1} \Lambda^{\pm} - b_A\Gamma_A^{\pm} -
  b_{A\pr} \Gamma_{A\pr}^{\pm}\ .
\end{equation}
Squaring both sides and using the fact that $\Gamma_A^{\pm}
\Gamma_{A\pr}^{\pm} = \pm \id$, which follows from
$\Gamma_A\Gamma_{A\pr} = \Gamma_C$, one gets
\begin{equation}
  g^{-2} = (a \pm b_C)^2 - (b_A \pm b_{A\pr})^2\ .
\end{equation}
The resonance condition is for $g$ to be real, \emph{viz.},
\begin{equation}
  \label{res-tbreak}
  |a\pm b_C| > |b_A \pm b_{A\pr}|
\end{equation}
if at least one of $\pm$ is satisfied. This is enumerated in the third
column in Table \ref{tab:tbreak} and plotted in
Fig.~\ref{tbreak-phase}.

As before, the expansion coefficients in Eq.~\ref{gz-tbreak} can be
estimated by the BCA and used to approximate the boundaries between the
resonant and non-resonant phases in the $(\omega,\eta)$ space. The
local Hamiltonian is $\hz_{00} = \alpha\id + \beta\Gamma^4 + \eta
\Gamma^{\mu\nu}$ where $\alpha = -\varepsilon_0$, $\beta = -12t\pr -
\lambda$. Its Green's function is $\bar G(\omega) = \bar a(\omega)\id
+ \bar b_1(\omega)\Gamma^4 + \bar b_2(\omega)\Gamma^{\mu\nu} + \bar
b_3(\omega) \Gamma^4 \Gamma^{\mu\nu}$ with
\begin{align}
  \notag
  \bar a(\omega) &= (\omega - \alpha) \left[(\omega - \alpha)^2 -
    \beta^2 - \eta^2\right]/Q(\omega) \\
  \notag
  \bar b_1(\omega) &= \beta \left[(\omega - \alpha)^2 - \beta^2 +
    \eta^2\right]/Q(\omega)\\
  \notag
  \bar b_2(\omega) &= \eta \left[(\omega -
    \alpha)^2 + \beta^2 - \eta^2\right]/Q(\omega)  \\
  \bar b_3(\omega) &= 2 (\omega - \alpha) \beta \eta/Q(\omega)
\end{align}
where $Q(\omega) = [(\omega - \alpha)^2 - (\beta + \eta)^2][(\omega -
\alpha)^2 - (\beta - \eta)^{2}]$. The resulting resonance conditions
are summarized in the fourth column of Table \ref{tab:tbreak}, and the
conditions for stable Weyl nodes (if exist) in the last column. 

The stable zones of the two classes $(+,-,-)$ (green in
Fig.~\ref{tbreak-phase}) and $(-,+,-)$ (red in
Fig.~\ref{tbreak-phase}) are restricted to opposite sides of a
critical value of the symmetry-breaking strength $\eta = \eta_c$.
Furthermore, near $\eta_c$, the region of stable energy narrows down
toward the Weyl node. One can think of the resonance energy as forming
an impurity band generated by a continuum of impurity strengths $g$.
Then the stable zones constitute gaps in such bands. In this sense,
$\eta_c$ marks a phase transition of the impurity band from gapless to
gapped. The existence of $\eta_c$ can be understood from the BCA,
according to which the phase boundaries are given by
\begin{equation}
  \omega = -\varepsilon_0 \pm \big(|\beta| - |\eta|\big)\ ,
\end{equation}
shown as gray dotted lines in Fig.~\ref{tbreak-phase}. These are the
two central band centers (eigenvalues of $\hz_{00}$). They cross at
$\eta = |\beta|$, which gives the critical strength $\eta_c$. $\eta_c
= 2$ in Fig.~\ref{tbreak-phase}. This is reminiscent of the bulk band
inversion in topological/Chern insulators that signifies a gapless to
gapped transition in their \emph{surface/edge} spectrum.

\begin{figure}
  \centering
  \begin{subfigure}[t]{0.44\textwidth}
    \includegraphics[width=\textwidth]{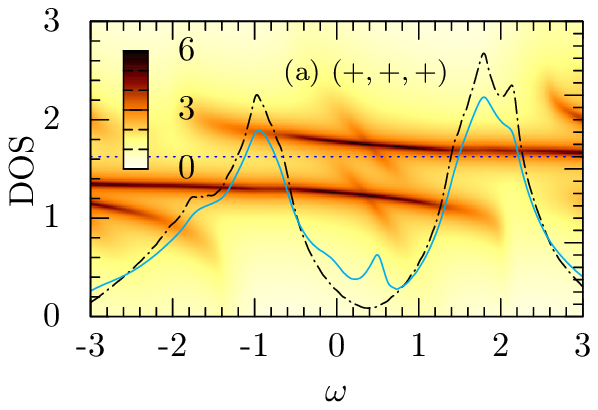}    
  \end{subfigure}
  \begin{subfigure}[t]{0.47\textwidth}
    \includegraphics[width=\textwidth]{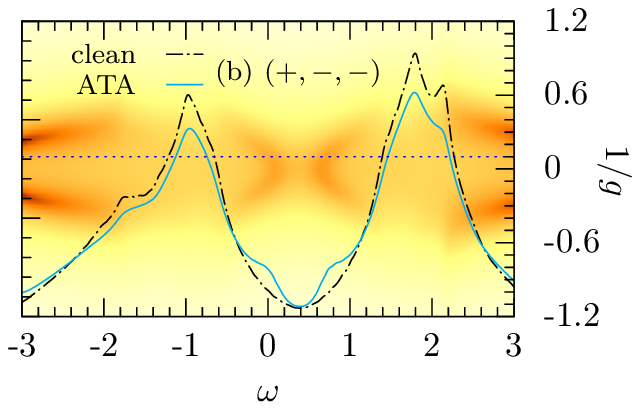}
  \end{subfigure}

  \begin{subfigure}[t]{0.44\textwidth}
    \includegraphics[width=\textwidth]{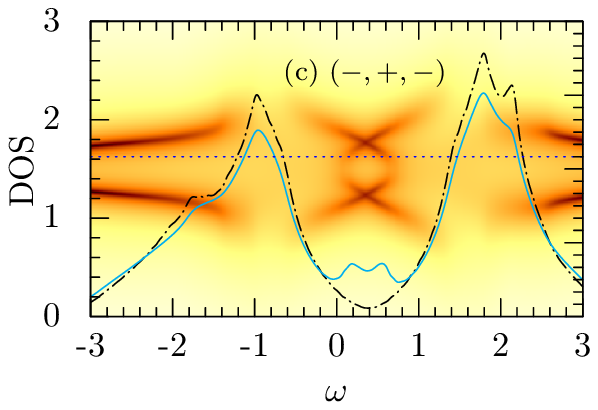}    
  \end{subfigure}
  \begin{subfigure}[t]{0.47\textwidth}
    \includegraphics[width=\textwidth]{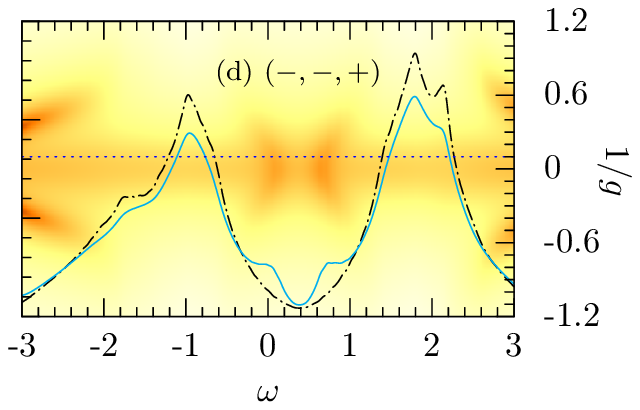}    
  \end{subfigure}
  \caption{(Color online) Effect of different classes of impurity
    ensembles on the ATA DOS. The color map shows $\log ||T_{00}(\omega +
    i\epsilon)||$, where darker color denotes larger $||T_{00}||$.
    Trails of darkest color will follow the zeros of $\ttr_{00}^{-1}$
    as given by the third column in Table ~\ref{tab:tbreak}. Black
    dash-dot curves are DOS of the clean system (identical in all four
    panels). Blue solid curves are DOS after adding impurity
    ensembles, and are computed using the average $T$-matrix
    approximation (ATA). Impurity class used in each panel are given
    in their respective caption. Impurity strengths are uniformly
    distributed in $g \in (0,10]$. Impurity concentration is $c =
    10\%$. The dotted horizontal line marks the minimum of $g^{-1}$
    ($0.1$): the ATA DOS is significantly enhanced for those $\omega$
    where the high $\log||T_{00}||$ lines exist \emph{above} this
    line. $\eta = 3$ is used; according to Fig.~\ref{tbreak-phase},
    the clean system is in the WS phase, and the DOS at the Weyl node
    should be suppressed for classes $(+,-,-)$ (b) and $(-,-,+)$ (d)
    but enhanced for the other two classes. This agrees with the plots
    shown here. Other parameters are the same as in
    Fig.~\ref{tbreak-phase}: $t = 0.05, t_1 = -0.5, t^{\prime} =
    -0.25, \lambda = 1, \varepsilon_0 = -0.3$, on a lattice of $N_x =
    N_y = N_z = 50$.  The spectral broadening (imaginary part of $\omega$) is taken
    to be $\epsilon = 0.05$, which prevents the DOS from touching zero (as it should) at the Weyl
    node energy near $\omega = 0.4$.}
  \label{tbreak-ata}
\end{figure}
To illustrate the above impurity band phase transition, we employ the
average $T$-matrix approximation (ATA) to investigate the
effect of an ensemble of local impurities spatially uniformly
distributed with concentration $c$ \cite{Economou06,balatsky06}. The entire ensemble has the same
matrix form, but with strength $g$ given by some distribution $f(g)$.  In the ATA formalism, statistical
averaging over the $f(g)$ will restore translational symmetry. The
impurity effect is then captured by a \emph{local} self energy,
$\Sigma_{\text{loc}} = c\langle T_{00}\rangle [1 + c\gz_{00}\langle
T_{00}\rangle]^{-1}$, where $\langle \cdots\rangle$ denotes the $f(g)$
averaging. The self-energy-corrected local Green's function is
$G_{\text{loc}}(z) = \frac{1}{N} \sum_{\mib k} 1/(z - \hz(\mib k) -
\Sigma_{\text{loc}})$ and the average LDOS is $\rho(\omega) = -\Im\Tr
\,G_{\text{loc}}(\omega + i0^{+})$. One then expect the ATA DOS,
$\rho_{\scriptscriptstyle{\rm ATA}}$, to be enhanced from the clean fraction,
$(1-c)\rho_{\text{clean}}$, for $\omega$ in the unstable phase but
reduced in the stable phase (since the integrated DOS is conserved).
This is shown in Fig.~\ref{tbreak-ata}, in which we plot at a fixed
$\eta$ the simplest case where $f(g)$ is a constant for $g \in (0,10]$
and zero otherwise. For this particular $\eta$ value, the Weyl nodal
energy is stable in (b) and (d), but is unstable in (a) and (c).

\section{Summary and discussion}
\label{sec:summary}
In this paper, we study the effect of localized impurities $V =
g\Lambda\delta(\mib x)$ on the bulk electronic structure of Weyl
semimetals. A general method is devised to detect whether or not a
resonance can be induced at energy $\omega$ by a $\Lambda$-type
impurity. If such a resonance is possible, then $\omega$ is said to be
unstable with respect to $\Lambda$, otherwise it is stable. The
stability of $\omega$ requires all eigenvalues of
$\grz_{00}(\omega)\Lambda$ to have finite imaginary part. Here,
$\grz_{00}(\omega)$ is the Hermitian part of the retarded local
Green's function $\gz_{00}(\omega + i0^{+})$. Otherwise, one can
always use a coupling strength $g = 1/u_a(\omega)$ to induce a
resonance at $\omega$, with $u_a(\omega)$ being the purely real
eigenvalue of $\grz_{00}(\omega) \Lambda$ indexed by $a$. The
existence of real $u_a(\omega)$ is equivalent to requiring the
Hermitian part of the inverse $T$ matrix to have a zero eigenvalue. An
immediate corollary is that impurities commuting with $\grz_{00}$ can
induce a resonance at an arbitrary energy, simply by tuning the impurity strength.
This includes the physically important class of local chemical potential perturbations.

We applied this method to four-band lattice Weyl semimetal models,
expressed in terms of Dirac $\Gamma$ matrices. For these models, the $T$
matrix and its eigenvalues can be obtained analytically.
Mathematically, one first classifies the clean Weyl semimetals
according to whether or not inversion ($\CI$\,) is broken. The
difference is in the decomposition of their local Green's functions
$\gz_{00}$. In each case, impurities are then classified by their
commutations with the $\Gamma$ matrices appearing in $\gz_{00}$:
anticommutation with components of $\gz_{00}$ result in a square-root
structure, which constrains the reality of $u_a(\omega)$. Note that
in this scheme, it is more relevant to know which $\Gamma$ matrices
appear than their exact numerical coefficients. For this purpose the
band center approximation (BCA) -- which replaces $\grz_{00}(\omega)$
by $(\omega - \hz_{00})^{-1}$ where $\hz_{00}$ is the local
Hamiltonian -- is quite useful as it has the same form of $\Gamma$
matrix decomposition with coefficients whose meanings are physically
more transparent. Results for $\CI$ breaking WS are reported in Table
\ref{tab:ibreak}, and for $\CI$ invariant WS in Table
\ref{tab:tbreak}. 

Realistic impurities are more likely to be linear combinations of
multiple $\Gamma$ matrices, mixing orbital, magnetic, and chemical
potential effects. A linear combination of impurities in the same
class, if it still squares to identity, is no different from a single
$\Gamma$ matrix in that class, and results obtained before hold
unchanged. While other combinations are not studied here, it is
reasonable to expect that stability will resemble that of the dominant
component if there is one, and crossover will happen as the relative
strengths change. We have confirmed this for several tractable cases
of Dirac semimetals. The method for obtaining the relation between
impurity strength $g$ and the induced resonance/bound state energy
$\omega$ may also prove useful in device engineering where specific
energy levels are desired. For Dirac materials with random strength
disorder, results similar to those shown in Fig.~\ref{tbreak-ata} are
expected, where roughly speaking impurity induced states form their
own band superimposed on the clean DOS, and stable energies constitute
the band gap. Such impurity bands may modify transport properties if
certain impurity ``superlattice'' is approximately formed, or if the
coherent length of single-impurity resonances become compatible with
the impurity density.

\section{Acknowledgement}
We are grateful to A.~Black-Schaffer, Tanmoy Das, and Da Wang for
useful discussions. This work was supported in part by the NSF through
grant DMR-1007028. Work at LANL was supported by US DoE Basic Energy
Sciences and in part by the Center for Integrated Nanotechnologies,
operated by LANS, LLC, for the National Nuclear Security
Administration of the U.S. Department of Energy under contract
DE-AC52-06NA25396. Work at Nordita was supported by ERC-DM321031 and
VR-621-2012-2983.

\newcommand{\hht}{{\wt H}} \newcommand{\oot}{{\wt \omega}}
\appendix
\section{Spectrum and Green's function of Eq.~\ref{hk0}}
\label{sec:app-g}
Here we derive the spectrum and Green's function of a generic Hamiltonian
\begin{equation}
  \label{hk0-app}
  H(\mib k) = \xi(\mib k) + \hht (\mib k)
\end{equation}
where
\begin{equation}
  \label{h1}
  \hht \equiv \sum_{a = 1}^5 d_a(\mib k)\Gamma^a + h \Gamma^{\mu\nu}\ .  
\end{equation}
Note that the index $a$ goes from $1$ to $5$, thus $d_4(\mib k)$ would
be $m(\mib k)$ in Eq.~\ref{hk0}. The symmetry breaking term is either
$\GSB = \Gamma^{\mu\nu}$, in which case its strength is $\eta =
h$, or $\GSB = \Gamma^a$ in which case $\eta = d_a$. In the
following, $\mib k$ dependence will be suppressed. From Eq.~\ref{h1},
it is easy to verify that
\begin{equation}
  \label{h2}
  \hht^2 = d^2 + h^2 + 2h\,\Gamma^{\mu\nu}\,\mib d_{\perp} \cdot
  \Gamma^{\perp}
\end{equation}
where $d^2 \equiv \sum_{a = 1}^5 d_a^2$ and $\mib d_{\perp}$ denotes
the three components ``perpendicular'' to the $\mu\nu$ ``plane'',
\emph{viz.}, $\mib d_{\perp}\cdot \Gamma^{\perp} = \sum_{a = 1}^5
d_a\Gamma^a - d_{\mu}\Gamma^{\mu} - d_{\nu}\Gamma^{\nu}$. The
``parallel'' components vanish in the cross term due to their
anticommutation with $\Gamma^{\mu\nu}$. Moving the scalars to the left
hand side and squaring again yields
\begin{equation}
  \label{denom-part}
  (\hht^2 - d^2 - h^2)^2 = 4h^2 d_{\perp}^2\ ,
\end{equation}
where we have used $[\Gamma^{\perp}, \Gamma^{\mu\nu}] = 0$.
Here $d_{\perp}$ is the magnitude of $\mib d_{\perp}$. Replacing $\hht$
with its eigenvalues $\wt E = E - \xi$ gives the spectrum of $H$,
\begin{equation}
  \label{hk0-eig}
  E = \xi \pm \sqrt{d^2 + h^2 \pm 2 h d_{\perp}}\ .
\end{equation}

The Green's function of Eq.~\ref{hk0-app} is (denoting $\oot = \omega
- \xi$)
\begin{align}
  G(\omega) &= \frac{1}{\omega - H} = \frac{1}{\oot - \hht} = \frac{\oot^2 + \hht}{\oot^2 - \hht^2} \\
  &= \frac{(\oot + \hht) (\oot - 2d^2 - 2h^2 + \hht^2)}{(\oot^2 - d^2 - h^2)^2 - (\hht^2 - d^2 - h^2)^2} \equiv \frac{M}{D}\ .
\end{align}
Using Eq.~\ref{denom-part}, the denominator $D$ is a number,
\begin{equation}
  \label{D}
  D = (\oot^2 - d^2 - h^2)^2 - 4h^2d_{\perp}^2
\end{equation}
which is nothing but $\prod_i (\omega - E_i)$ with $E_i$ given by
Eq.~\ref{hk0-eig}. The numerator in powers of $\hht$ is
\begin{equation}
  \label{numerator}
  M = \oot(\oot^2 - 2d^2 - 2h^2) + (\oot^2 - 2d^2 - 2h^2) \hht + \oot
  \hht^2 + \hht^3,
\end{equation}
in which $\hht^2$ is already given by Eq.~\ref{h2}, and
\begin{equation}
  \hht^3 = \hht \hht^2 = (d^2 + h^2) \hht
  + 2h\,\mib d_{\perp}\cdot\Gamma^{\perp} \,\sum_ad_a\Gamma^a \,\Gamma^{\mu\nu}
  + 2h^2 \mib d_{\perp}\cdot \Gamma^{\perp}\ .
\end{equation}
Rewriting $\sum_a d_a\Gamma^a = \mib d_{\perp}\cdot \Gamma^{\perp} +
d_{\mu}\Gamma^{\mu} + d_{\nu}\Gamma^{\nu}$, and using
$\Gamma^{\mu(\nu)}\Gamma^{\mu\nu} = +(-) i\Gamma^{\nu(\mu)}$, we have
\begin{equation}
  \label{h3}
  \hht^3 = (d^2 + h^2) \hht + 2h d_{\perp}^2\,\Gamma^{\mu\nu} +
  2h(h + id_{\mu}\Gamma^{\nu} - id_{\nu}\Gamma^{\mu})\,\mib d_{\perp} \cdot
  \Gamma^{\perp}\ .
\end{equation}
Substituting Eqs.~\ref{h1}, \ref{h2} and \ref{h3} in Eq.~\ref{numerator} gives
\begin{multline}
  \label{M}
  M = \oot(\oot^2 - d^2 - h^2) + (\oot^2 - d^2 - h^2)\ \hht +
  2hd_{\perp}^2\ \Gamma^{\mu\nu} \\
  + 2h\Bigl[\oot \Gamma^{\mu\nu} + h
    + i d_{\mu}\Gamma^{\nu} - i d_{\nu}\Gamma^{\mu}\Bigr]\ \mib
  d_{\perp}\cdot \Gamma^{\perp}\ .
\end{multline}

Eqs.~\ref{D} and \ref{M} can now be used to obtain the local Green's
functions. Note that since $d_i(-\mib k) = -d_i(\mib k)$ for $i =
1,2,3$ in the Hamiltonian of Eq.~\ref{hk0}, many terms in Eq.~\ref{M}
will vanish upon $\mib k$-space averaging.
\vspace{2mm}\\
(1) If $\GSB = \Gamma^{\mu 4}$, $\mu \neq 4$ (see
Sec.~\ref{sec:ibreak}), then we have $\mib d = (d_1,d_2, d_3, m, 0)$
and $h = \eta$. Upon ${\mib k}$-space averaging, denoted by $\langle \cdots \rangle$, $\hht \rightarrow
m\Gamma^4 + \eta \Gamma^{\mu 4}$ and $\mib d_{\perp}\cdot
{\mib\Gamma}^{\perp} \rightarrow 0$ in Eq.~\ref{M}, yielding
\begin{equation}
  \label{gz00-gm4}
  \gz_{00}(\omega) = a(\omega) + b_1(\omega)\Gamma^4 + b_2(\omega) \Gamma^{\mu 4}\ ,
\end{equation}
with
\begin{align}
  \notag
  a(\omega) &= \left< \frac{\oot(\oot^2 - d^2 - \eta^2)}{(\oot^2 - d^2 - \eta^2)^2 - 4\eta^2d_{\perp}^2}\right>
  =  \frac{1}{2} \left< \frac{\wt \omega}{\wt \omega^2 - \wt E_+^2} + \frac{\wt \omega}{\wt \omega^2 - \wt E_-^2}\right>\\
  \notag
  b_1(\omega) &= \left< \frac{m(\oot^2 - d^2 - \eta^2)}{(\oot^2 - d^2 - \eta^2)^2 - 4\eta^2d_{\perp}^2}\right>
  = \frac{1}{2}\left<
      \frac{m}{\wt \omega^2 - \wt E_+^2} + \frac{m}{\wt \omega^2 - \wt E_-^2}
    \right>
    \\
  \notag
  b_2(\omega) &= \eta\left< \frac{\oot^2 - d^2 - \eta^2 +
      2d_{\perp}^2}{(\oot^2 - d^2 - \eta^2)^2 -
      4\eta^2d_{\perp}^2}\right>
  \\
  \label{coef:gmu4}
  &= \frac{\eta}{2} \left<
    \frac{1}{\wt \omega^2 - \wt E_+^2} + \frac{1}{\wt \omega^2 - \wt E_-^2}
    + \frac{4 d_{\perp}^2}{(\wt \omega^2 - \wt E_+^2)(\wt\omega^2 - \wt E_-^2)}
    \right>
\end{align}
where $\wt E_{\pm}^2 = (E-\xi)^2 = d_{\parallel}^2 + (d_{\perp} \pm
\eta)^2$, and
\begin{equation}
  \oot = \omega - \xi(\mib k)\quad,\quad\\
  d^2 = \sum_{i=1}^3d_i(\mib k)^2 + m(\mib k)^2\quad,\quad\\
  \label{coef:gmu4-end}
  d_{\perp}^2 = \sum_{i=1}^3d_i(\mib k)^2(1 - \delta_{\mu,i})\ .
\end{equation}
\vspace{2mm}\\
(2) if $\GSB = \Gamma^{\mu}$, $\mu \neq 4$, (see
Sec.~\ref{sec:ibreak}), then $\mib d = \eta \,\mib e_{\mu} +
(d_1,d_2,d_3,m,0)$ and $h = 0$. Upon $\mib k$\,-averaging, Eq.~\ref{M}
is effectively $M = \oot(\oot^2 - d^2) + (\oot^2 - d^2)(m\Gamma^4 +
\eta\Gamma^{\mu})$, thus
\begin{equation}
  \gz_{00}(\omega) = a(\omega) + b_1(\omega) \Gamma^4 + b_2(\omega) \Gamma^{\mu}
\end{equation}
where
\begin{equation}
  \label{coef:gmu}
  a(\omega) = \left\langle\frac{\oot}{\oot^2 - d^2}\right\rangle\quad,\quad
  b_1(\omega) = \left\langle\frac{m}{\oot^2 - d^2}\right\rangle\quad,\quad
  b_2(\omega) = \eta\left\langle\frac{1}{\oot^2 - d^2}\right\rangle
\end{equation}
with
\begin{equation}
  \oot = \omega - \xi(\mib k)\ , \\
  \label{coef:gmu-end}
  d^2 = \sum_{i=1}^3d_i(\mib k)^2 + m^2 + \eta^2\ .
\end{equation}
\vspace{2mm}\\
(3) If $\GSB = \Gamma^{\mu\nu}$, $\mu \neq \nu \neq 4$ (see
Sec.~\ref{sec:tbreak}), then $\mib d = (d_1, d_2, d_3, m, 0)$ and $h =
\eta$. Upon ${\mib k}$-space average, $\hht \rightarrow m\Gamma^4 + \eta
\Gamma^{\mu\nu}$, $\mib d_{\perp}\cdot \Gamma^{\perp} = m\Gamma^4$,
$d_{\mu}\Gamma^{\nu}$ and $d_{\nu}\Gamma^{\mu}\rightarrow 0$, thus
\begin{equation}
  \gz_{00}(\omega) = a(\omega) + b_1(\omega)\Gamma^4 + b_2(\omega)\Gamma^{\mu\nu} + b_3(\omega)\Gamma^4\Gamma^{\mu\nu}
\end{equation}
where
\begin{align}
  \notag
  a(\omega) &= \left< \frac{\oot(\oot^2 - d^2 - \eta^2)}{(\oot^2 - d^2 - \eta^2)^2 - 4\eta^2d_{\perp}^2}\right>
  = \frac{1}{2} \left<
    \frac{\wt \omega}{\wt \omega^2 - \wt E_+^2} + \frac{\wt \omega}{\wt \omega^2 - \wt E_-^2}
  \right>
  \\
  \notag
  b_1(\omega) &= \left< \frac{m(\oot^2 - d^2 + \eta^2)}{(\oot^2 - d^2 - \eta^2)^2 - 4\eta^2d_{\perp}^2}\right>
  = \frac{1}{2} \left<
      \frac{m}{\wt \omega^2 - \wt E_+^2} + \frac{m}{\wt \omega^2 - \wt E_-^2} + \frac{4m\eta^2}{(\wt \omega^2 - \wt E_+^2)(\wt \omega^2 - \wt E_-^2)}
    \right>
  \\  
  \notag
  b_2(\omega) &= \eta\left< \frac{\oot^2 - d^2 - \eta^2 +
      2d_{\perp}^2}{(\oot^2 - d^2 - \eta^2)^2 -
      4\eta^2d_{\perp}^2}\right>
  = \frac{\eta}{2} \left<
    \frac{1}{\wt \omega^2 - \wt E_+^2} + \frac{1}{\wt \omega^2 - \wt E_-^2} + \frac{4d_{\perp}^2}{(\wt \omega^2 - \wt E_+^2)(\wt \omega^2 - \wt E_-^2) }
  \right>
  \\
  \label{coef:gmunu}
  b_3(\omega) &= 2\eta \left< \frac{\oot m}{(\oot^2 - d^2 - \eta^2)^2 - 4\eta^2d_{\perp}^2} \right>
  = 2\eta\left<
    \frac{\wt \omega m}{(\wt \omega^2 - \wt E_+^2)(\wt \omega^2 - \wt E_-^2) }
  \right>
\end{align}
with $\wt E_{\pm}^2 = (E-\xi)^2 = d_{\parallel}^2 + (d_{\perp} \pm
\eta)^2$, and
\begin{equation}
  \oot = \omega - \xi(\mib k)\ ,\\
  d^2 = \sum_{i=1}^3 d_i(\mib k)^2 + m(\mib k)^{2}\ ,
\end{equation}
and
\begin{equation}
  \label{coef:gmunu-end}
  d_{\perp}^2 = \sum_{i=1}^3 d_i(\mib k)^2(1 - \delta_{\mu,i} - \delta_{\nu,i})
  + m(\mib k)^2(1 - \delta_{\mu,4} - \delta_{\nu,4})\ .
\end{equation}

\section{Green's function of the BHB theory with prototypical $\GSB$}
\label{sec:app-cont}
In this section we use the following convention for $\Gamma$ matrices,
\begin{gather}
  \Gamma^1 = \id_{\tau} \otimes \sigma_x \quad , \quad
  \Gamma^2 = \id_{\tau} \otimes \sigma_y \quad , \quad
  \Gamma^3 = \tau_x \otimes \sigma_z \quad , \quad
  \Gamma^4 = \tau_y \otimes \sigma_z \quad , \quad
  \Gamma^5 = \tau_z \otimes \sigma_z \ ,
\end{gather}
which is related to the one used in the text by a unitary
transformation. 
\subsection{WS with point nodes ($\GSB = \Gamma^{21}$)}
\label{sec:app-cont-g21}
The unperturbed Hamiltonian is
\begin{gather}
  H(\vec k) = \sum_{i=1}^3 d_i(k_i) \Gamma^i + m \Gamma^4 + \eta \Gamma^{21}
\end{gather}
Taking $d_i(k_i) = k_i$ will give the linearized BHB Hamiltonian.
Explicitly,
\begin{gather}
  H(\Bk) =  \id_{\tau} \otimes h_{\sigma}  + h_{\tau} \otimes \sigma_z \ ,\\
  h_{\sigma}(k_x, k_y) = d_1(k_x) \sigma_x + d_2(k_y) \sigma_y\quad ,
  \quad h_{\tau}(k_z) = d_3(k_z) \tau_x + m \tau_y + \eta \id_{\tau}\ .
\end{gather}
Diagonalizing $h_{\tau}$ brings $H$ into block-diagonal form,
\begin{gather}
  U\dg h_{\tau} U = \eta\id_{\tau} + \sqrt{d_3^2 + m^2}\, \tau_z \ ,\\
  \wt H \equiv \mathcal{U}\dg H \mathcal{U} =
  \begin{pmatrix}
    H_+ \\ & H_-
  \end{pmatrix}
\end{gather}
where $U(k_z)$ and $\mathcal{U}(k_z)$ are unitary matrices acting on
the $\tau$ space and the $\tau \otimes \sigma$ space, respectively,
\begin{gather}
  U(k_z) = \exp \left(-\frac{i\phi(k_z)}{2} \tau_z\right) \exp\left( -\frac{i\pi}{4} \tau_y\right)\ , \\
  \mathcal{U}(k_z) = U(k_z) \otimes \id_{\sigma}\ ,\\
  \phi(k_z) = \tan^{-1} \frac{m}{d_3(k_z)}\ ,
\end{gather}
and the diagonal blocks of $\wt H$ are labeled by $\tau = \pm 1$ with
\begin{gather}
  H_\tau = \vec B^{\tau} \cdot \vec \sigma\quad , \quad \vec
  B^{\tau} = (d_1\ , \ d_2\ , \ \eta + \tau \sqrt{d_3^2 + m^2})\quad ,
  \quad \tau = \pm 1\ .
\end{gather}
The eigenvalues of $H$ are thus $\pm E_{\tau}$,
\begin{gather}
  \label{erg-tau}
  E_{\tau} = |\vec B^{\tau}| = \sqrt{d^2 + m^2 + \eta^2 + 2 \tau\eta \sqrt{d_3^2 + m^2}}\ .
\end{gather}
Note that for nonzero $\eta$ and $m$, bands with different $\tau$
indices can never cross. 

Weyl nodes only exist in the $\tau = -1$ subspace in which the two
bands touch at
\begin{gather}
  \vec d = (0\ , \ 0 \ , \ \pm\Delta)\quad , \quad \Delta = \sqrt{\eta^2 - m^2}\ .
\end{gather}
In the vicinity of $d_3 = \pm \Delta$, one writes
\begin{gather}
  d_3 = c(\Delta + q)\quad , \quad c = \pm 1\ ,
\end{gather}
then for $q \ll \eta$,
\begin{gather}
  H_-(\vec k) = d_1\, \sigma_x + d_2\, \sigma_y - \Delta \frac{q}{\eta}\,\sigma_z + {\cal O}(q^2/\eta^2)\ ,
\end{gather}
and its spectrum is
\begin{gather}
  \lambda = \pm \sqrt{d_1^2 + d_2^2 + q^2 \frac{\Delta^2}{\eta^2}}\ .
\end{gather}
Note that $\Delta / \eta$ is related to the $\phi$ angle of the Weyl
nodes via
\begin{gather}
   \cos \phi_c = c \frac{\Delta}{\eta}\quad , \quad \phi_c \equiv \phi|_{d_3 = c \Delta}\ .
\end{gather}

The local Green's function is
\begin{gather}
  \label{gloc-g12}
  \gz_{00}(\omega) = \left< \mathcal{U}(k_z)
  \begin{pmatrix}
    G_+(\omega, \vec k) \\ & G_-(\omega, \vec k)
  \end{pmatrix}
  \mathcal{U}\dg(k_z)\right>
\end{gather}
where $\langle\cdots\rangle$ denotes $\vec k$-space average, and
\begin{gather}
  G_{\tau}(\omega, \vec k) = \frac{\omega\id_{\sigma} +
    B^{\tau}_z(k_z) \sigma_z}{ \omega^2 - |\vec B^{\tau}(\vec
    k)|^2}\quad , \quad \tau = \pm 1
\end{gather}
$G_{\tau}$ is obtained from the Green's function of $H_{\tau}(\vec k)$
by dropping terms odd in $k_x$ and $k_y$ which would have averaged to
zero.

Now we turn to the linearized theory $d_i(k_i) = k_i$. The aim is to
isolate the contribution to the impurity effect near the nodal energy
$\omega \sim 0$, from states near the Weyl nodes. The following
approximations will be made:
\begin{enumerate}
\item We reduce the full $\vec k$-space to two isotropical spheres of
  radius $Q$ around the two Weyl nodes labeled by their chirality $c =
  \pm 1$: $\vec k = (k_x, k_y, c(\Delta + q))$ for $k_x, k_y, q \in
  [-Q, Q]$. In other words, the $\vec k$-space average $\int d^3k
  \rightarrow \sum_{c=\pm 1}\int dk_x dk_y dq$.
\item Within these spheres we will approximate $\mathcal{U}(k_z)$ by
  $\mathcal{U}(c\Delta)$, \emph{i.e.} its value on the nodes, which is
  then moved out of $\langle \cdots \rangle$ in eq.~\ref{gloc-g12}.
\item Further more, since the $\tau = 1$ eigenstates are gapped,
  $G_-(\omega, \vec k) \gg G_+(\omega, \vec k)$ so in
  eq.~\ref{gloc-g12} one can set $G_+ = 0$, \emph{i.e.}, project onto
  the $\tau = -1$ subspace.
\end{enumerate}
Under these approximations the local Green's function
becomes
\begin{align}
  \gz_{00}(\omega) &= \left[\sum_{c = \pm 1} U(c\Delta)
  \begin{pmatrix}
    0 \\ & 1
  \end{pmatrix} U\dg(c\Delta)\right] \otimes \langle G_-(\omega, \vec k)\rangle\\
\label{gz00-g12-linear}
&= \gz_-(\omega) \ \left( \id_{\tau} - \sin \phi_+ \tau_y\right) \otimes \id_{\sigma}\ ,
\end{align}
where $\phi_+$ is the $\phi$ angle on the positive chirality node,
\begin{gather}
  \sin \phi_+ = \frac{m}{\eta}
\end{gather}
and
\begin{gather}
  \label{g12-gscalar}
  \gz_-(\omega) \equiv \langle G_-(\omega, \vec k)\rangle =
  \int\limits^Q \frac{d k_x d k_y dq}{(2\pi)^3} \frac{\omega - q
    \frac{\Delta}{\eta} \sigma_z}{\omega^2 - k_x^2 - k_y^2 - q^2
    \frac{\Delta^2}{\eta^2}}\ .
\end{gather}
Note that $\langle G_-(\omega, \vec k)\rangle$ is proportional to
$\id_{\sigma}$ because the coefficient of $\sigma_z$ is odd and
integrates to zero. Introducing
\begin{gather}
  x = \frac{m}{\eta}\cos \theta\quad , \quad u(x) = 1-x^2 \quad , \quad \kappa = \sqrt{u} \sqrt{k_x^2 + k_y^2 + q^2}\ ,
\end{gather}
and using
\begin{gather}
  \int\limits_0^K \frac{\kappa^2 d\kappa}{\omega^2 - \kappa^2} = -K + \frac{\omega}{2} \log \frac{\omega + K}{\omega - K}
\end{gather}
one has 
\begin{align}
  \gz_-(\omega) &= \frac{\omega}{2\pi^2}
  \frac{\eta}{m}\int\limits_0^{ \frac{m}{\eta}} \frac{dx}{u \sqrt{u}}
  \int\limits_0^{Q \sqrt{u}} \frac{\kappa^2 d\kappa}{\omega^2 -
    \kappa^2}\\
  &= \frac{Q\omega}{4\pi^2 \sin \phi_+} \log \frac{1 - \sin \phi_+}{ 1 + \sin\phi_+}
  + \frac{\omega^2}{4\pi^2 \sin\phi_+} \int\limits_0^{\phi_+} d(\tan \phi)\, \log \frac{\omega + Q \cos \phi}{\omega - Q\cos \phi}\ ,
\end{align}
where we have used $\sin \phi_+ = m/\eta$ and introduced $\phi =
\sin^{-1} x$. In the limit $|\omega| \ll Q$, the second integral
becomes $-i\pi\tan \phi_+$ (using $\omega \rightarrow \omega +
i0^+$).

\subsection{WS with nodal line ($\GSB = \Gamma^{35}$)}
\label{sec:app-g35}
Consider the Hamiltonian
\begin{align}
  H(\vec k) &= \sum_{i=1}^3 d_i(k_i) \Gamma^i + m\Gamma^4 + \eta
  \Gamma^{35}\\
  &= \id_{\tau} \otimes (d_1 \sigma_x + d_2 \sigma_y) +
  \tau_y \otimes (\eta \id_{\sigma} + m \sigma_z) + d_3 \tau_x \otimes \sigma_z\ .
\end{align}
To block diagonalize, we first rotate $(\tau_y, \tau_z) \rightarrow
(\tau_z, -\tau_y)$, and then send, simultaneously, $\tau_{\pm} \otimes
\id_{\sigma} \rightarrow \tau_{\pm} \otimes \sigma_z$ and $\id_{\tau}
\otimes \sigma_{\pm} \rightarrow \tau_z \otimes \sigma_{\pm}$, which
is the unitary transformation $U = \textsf{diag}(1,1,1,-1)$. This is
equivalent to taking the following $\Gamma$ matrix convention from the
outset (switching the order of $\tau$ and $\sigma$ spaces in the
direct product),
\begin{gather}
  \Gamma^1 = \sigma_x \otimes \tau_z\quad , \quad \Gamma^2 = \sigma_y
  \otimes \tau_z \quad , \quad \Gamma^3 = \id_{\sigma} \otimes
  \tau_x\quad , \quad \Gamma^4 = \sigma_z \otimes \tau_z\quad , \quad
  \Gamma^5 = - \id_{\sigma} \otimes \tau_y\ .
\end{gather}
After this basis change, one has
\begin{gather}
  H(\vec k) = h_{\sigma} \otimes \tau_z + d_3 \id_{\sigma} \otimes
  \tau_x \quad , \quad h_{\sigma} = d_1 \sigma_x + d_2 \sigma_y + m
  \sigma_z + \eta \id_{\sigma}\ .
\end{gather}
Diagonalizing $h_{\sigma}$ then brings $H(\vec k)$ into block-diagonal
form,
\begin{gather}
  U\dg h_{\sigma} U = \eta \id_{\sigma} + \sqrt{d_1^2 + d_2^2 + m^2}\> \sigma_z\ , \\
  \wt H \equiv \mathcal{U}\dg H \mathcal{U} =
  \begin{pmatrix}
    H_+ \\ & H_-
  \end{pmatrix}
\end{gather}
where $U(k_x, k_y)$ and $\mathcal{U}(k_x, k_y)$ are unitary matrices
acting on the $\sigma$ space and the $\sigma\otimes \tau$ space,
respectively,
\begin{gather}
  U(k_x, k_y) = \exp\left(
    -\frac{i}{2}\phi(k_x, k_y)\sigma_z
  \right) \exp\!\left(
    -\frac{i}{2} \theta(k_x, k_y)\sigma_y
  \right)\ ,\\
  \mathcal{U}(k_x, k_y) = U(k_x, k_y) \otimes \id_{\tau}\ ,
\end{gather}
and $\theta$, $\phi$ are the polar and azimuthal angles of the
vector $(d_1, d_2, m)$. The diagonal blocks of $\wt H$ are labeled by
$s = \pm 1$ with
\begin{gather}
  H_s = d_3\, \tau_x + \left(\eta + s \sqrt{d_1^2 + d_2^2 + m^2}\right) \tau_z\quad , \quad s = \pm 1.
\end{gather}
The eigenvalues of $H$ are thus $\pm E_s$,
\begin{gather}
  E_s = \sqrt{ d_3^2 + \left(\eta + s \sqrt{d_1^2 + d_2^2 + m^2}\right)^{\!\!2}}\ .
\end{gather}
Note that for nonzero $\eta$ and $m$, bands with different $s$ indices
can never cross.

Weyl nodes only exist in the $s = -1$ subspace in which the two bands
touch at
\begin{gather}
  \vec d = (\Delta \cos \phi\ , \Delta \sin \phi\ , \ 0) \quad , \quad
  \Delta = \sqrt{\eta^2 - m^2}\ .
\end{gather}
In the vicinity of the line node, one writes
\begin{gather}
  d_1 = (\Delta + q) \cos\phi\quad , \quad d_2 = (\Delta + q)\sin\phi\ ,
\end{gather}
then for $q \ll \eta$,
\begin{gather}
  H_-(\vec k) = d_3\, \tau_x - \Delta\, \frac{q}{\eta} \,\tau_z + {\cal O}(q^2/\eta^2)\ ,
\end{gather}
and its spectrum is
\begin{gather}
  \lambda = \pm \sqrt{d_3^2 + q^2 \frac{\Delta^2}{\eta^2}}\ .
\end{gather}
Note that $\Delta / \eta$ is related to the polar angle $\theta$ of
the vector $(d_1, d_2, m)$ on the nodal line,
\begin{gather}
  \sin \theta_N = \frac{\Delta}{\eta}
\end{gather}

The local Green's function is
\begin{gather}
  \label{gloc-g35}
  \gz_{00}(\omega) = \left<
    \mathcal{U}(k_x, k_y)
    \begin{pmatrix}
      G_+(\omega, \vec k) \\ & G_-(\omega, \vec k)
    \end{pmatrix}
    \mathcal{U}\dg(k_x, k_y)
  \right>
\end{gather}
where $\langle \cdots \rangle$ denotes $\vec k$-space average, and
\begin{gather}
  G_s(\omega, \vec k) = \frac{\omega \id_{\tau} + \left(\eta + s
      \sqrt{d_1^2 + d_2^2 + m^2}\right) \tau_z}{\omega^2 - d_3^2 -
    \left(\eta + s \sqrt{d_1^2 + d_2^2 + m^2}\right)^2}\quad , \quad s = \pm 1\ .
\end{gather}
$G_s$ is obtained from the Green's function of $H_s(\vec k)$ by
dropping terms odd in $k_z$, \emph{i.e.} the one proportional to
$\tau_x$ in the numerator, which would have averaged to zero.

Now we turn to the linearized theory $d_i = k_i$ and investigate the
contribution of states near the line Weyl node to the impurity effect
near zero energy $\omega \sim 0$. We employ the following
approximations,
\begin{enumerate}
\item The full $\vec k$-space is reduced to a tube of radius $Q$
  around the line node, $\vec k = ((\Delta + q) \cos \phi, (\Delta +
  q) \sin \phi, k_z)$ with $\sqrt{q^2 + k_z^2} \in [0, Q]$. The $\vec
  k$-space average $\int d^3 k \rightarrow (\int d\phi/2\pi)(\int 2\pi \Delta\, dq\, dk_z)$.
\item Within the tube we will approximate $\mathcal{U}(k_x, k_y)$ by
  its value on the nodal line, $\mathcal{U}(\Delta\cos \phi, \Delta
  \sin\phi)$. It is then taken out of the average over the tube's
  cross-section.
\item Since the $s = 1$ states are gapped, $G_-(\omega, \vec k) \gg
  G_+(\omega, \vec k)$ for $\omega \ll \eta$, so in eq.~\ref{gloc-g35}
  one can set $G_+ = 0$, \emph{i.e.}, project onto the $s = -1$
  subspace.
\end{enumerate}
Under these approximations, the local Green's function becomes
\begin{align}
  \gz_{00} &= \left< U(\Delta, \phi)
    \begin{pmatrix}
      0 \\ & 1
    \end{pmatrix} U\dg(\Delta, \phi) \right>_{\phi} \otimes \langle
  G_-(\omega, \vec k)\rangle_{q, k_z}\\
  \label{gz00-g35-linear}
  &=  \frac{\Delta}{2} \gz_-(\omega) (\id_{\sigma} - \cos \theta_N \sigma_z) \otimes \id_{\tau}\ ,
\end{align}
where $\theta_N$ is the aforementioned polar angle of the nodal line
in the $(k_x, k_y, m)$ space,
\begin{gather}
  \cos \theta_N = \frac{m}{\eta}
\end{gather}
and
\begin{gather}
  \gz_-(\omega) \equiv \langle G_-(\omega, \vec k)\rangle_{q,k_z} =
  \int\limits^Q \frac{dq\,dk_z}{(2\pi)^2}
  \frac{\omega - \Delta\, \eta^{-1}q\,\sigma_z}{\omega^2 - k_z^2 - \Delta^2\eta^{-2} q^2}\ .
\end{gather}
Note that $\gz_-(\omega)$ is a number because the coefficient of
$\sigma_z$ is odd in $q$ and integrates to zero. Denoting
\begin{gather}
  \tan \chi = \frac{q}{k_z}\quad , \quad u(\chi) = 1 -
  \frac{m^2}{\eta^2}\sin^2\chi \quad , \quad \kappa =
  \sqrt{u}\sqrt{k_z^2 + q^2}\ ,
\end{gather}
one has
\begin{align}
  \gz_-(\omega) &= \frac{\omega}{4\pi^2} \int\limits_0^{2\pi}
  \frac{d\chi}{u}\int\limits_0^{Q \sqrt{u}} \frac{\kappa
    d\kappa}{\omega^2 - \kappa^2}\\
  &= -\frac{\omega}{2\pi^2} \int\limits_0^{\pi/2} \frac{d\chi}{u} \log \left[ 1 - u \frac{Q^2}{\omega^2}\right]\ .
\end{align}
In the limit $\omega \ll Q$,
\begin{gather}
  \log(1 - uQ^2/\omega^2) \simeq \log\left[- uQ^2/(\omega+i0^+)^2\right] =
  \log(uQ^2/\omega^2) + i\pi \textsf{sgn}(\omega)\ ,
\end{gather}
thus
\begin{align}
  \gz_-(\omega) &= \Biggl(\frac{Q}{\pi^2}\cdot \overbrace{\frac{\omega}{Q}\log \frac{|\omega|}{Q}}^{\mathclap{x \log x \rightarrow 0 \text{ for } x \rightarrow 0}}  - i \frac{|\omega|}{2\pi}\Biggr) \int\limits_0^{\pi/2} \frac{d\chi}{u(\chi)}
  - \frac{\omega}{2\pi^2} \int\limits_0^{\pi/2} d\chi \frac{\log u(\chi)}{u(\chi)}\\
  &= R(\theta_N)\, \omega - i \frac{|\omega|}{4\sin \theta_N}\ , 
\end{align}
with
\begin{gather}
  \label{rint}
  R(\theta_N) = - \int\limits_0^{\pi/2} \frac{d\chi}{2\pi^2} \frac{\log (1 - \cos^2 \theta_N \sin^2\chi)}{1-\cos^2\theta_N \sin^2\chi}\ .
\end{gather}

\section*{References}
\bibliographystyle{h-physrev}
\bibliography{weyl}
\end{document}